# Privacy-Preserving Multi-center Differential Protein Abundance Analysis with FedProt

Yuliya Burankova[1,2], Miriam Abele[1,3], Mohammad Bakhtiari[2], Christine von Törne[4], Teresa Barth[5], Lisa Schweizer[6], Pieter Giesbertz[7], Johannes R. Schmidt[8], Stefan Kalkhof[8,9], Janina Müller-Deile[10], Peter A van Veelen[11], Yassene Mohammed[11], Elke Hammer[12,13], Lis Arend[2,14], Klaudia Adamowicz[2], Tanja Laske[2], Anne Hartebrodt[15,16], Tobias Frisch[15], Chen Meng[3], Julian Matschinske[2], Julian Späth[2], Richard Röttger[15], Veit Schwämmle[15], Stefanie M. Hauck[4], Stefan Lichtenthaler[7,17,18], Axel Imhof[5], Matthias Mann[6], Christina Ludwig[3], Bernhard Kuster[1], Jan Baumbach[2,15,*], Olga Zolotareva[2,14,*]

[1]Chair of Proteomics and Bioanalytics, TUM School of Life Sciences, Technical University of Munich, Freising, Germany
[2]Institute for Computational Systems Biology, University of Hamburg, Hamburg, Germany
[3]Bavarian Center for Biomolecular Mass Spectrometry (BayBioMS), TUM School of Life Sciences, Technical University of Munich, Freising, Germany
[4]Metabolomics and Proteomics Core, Helmholtz Center Munich, Munich, Germany
[5]Protein Analysis Unit, Biomedical Center, Faculty of Medicine, LMU Munich, Martinsried, Germany
[6]Max Planck Institute of Biochemistry, Martinsried, Germany
[7]German Center for Neurodegenerative Diseases (DZNE), Munich, Germany
[8]Department of Preclinical Development and Validation, Fraunhofer Institute for Cell Therapy and Immunology IZI, Leipzig, Germany
[9]Institute for Bioanalysis, University of Applied Science Coburg, Coburg, Germany
[10]Department of Nephrology, University Hospital Erlangen, Erlangen, Germany
[11]Center for Proteomics and Metabolomics, Leiden University Medical Center, Leiden, The Netherlands
[12]University of Greifswald, Greifswald, Germany
[13]German Center for Cardiovascular Diseases (DZHK), Germany
[14]Data Science in Systems Biology, TUM School of Life Sciences, Technical University of Munich, Freising, Germany
[15]University of Southern Denmark, Denmark
[16]Friedrich-Alexander-Universität Erlangen-Nürnberg, Erlangen, Germany
[17]Neuroproteomics, School of Medicine and Health, Klinikum rechts der Isar, Technical University of Munich, Munich, Germany
[18]Munich Cluster for Systems Neurology (SyNergy), Munich, Germany

*Joint last authors

# Abstract

Quantitative mass spectrometry has revolutionized proteomics by enabling simultaneous quantification of thousands of proteins. Pooling patient-derived data from multiple institutions enhances statistical power but raises significant privacy concerns. Here we introduce FedProt, the first privacy-preserving tool for collaborative differential protein abundance analysis of distributed data, which utilizes federated learning and additive secret sharing. In the absence of a multicenter patient-derived dataset for evaluation, we created two, one at five centers from LFQ *E.coli* experiments and one at three centers from TMT human serum. Evaluations using these datasets confirm that FedProt achieves accuracy equivalent to *DEqMS* applied to pooled data, with completely negligible absolute differences no greater than $4*10^{-12}$. In contrast, $-\log_{10}$(p-values) computed by the most accurate meta-analysis methods diverged from the centralized analysis results by up to 25-27. FedProt is available as a web tool with detailed documentation as a FeatureCloud App.

# Introduction

The expansion of proteomics data is an invaluable resource, unlocking significant potential for large-scale biomedical research. While genomics provides a static view of an organism's potential capabilities, mass spectrometry (MS)-based proteomics offers a comprehensive insight into the dynamic protein composition, interactions, and modifications that cannot be inferred solely from genomics or transcriptomics data (Aebersold and Mann, 2016; Altelaar et al., 2013). The MS-based proteomics enhances our understanding of the proteome's dynamic nature, composition, structure, and function.

Techniques such as data-independent acquisition (DIA) MS have allowed simultaneous quantification of thousands of proteins (Muntel et al., 2019) with wide proteome coverage and low missing values (Bruderer et al., 2017; Fröhlich et al., 2024). Functionally, DIA achieves this by systematically fragmenting all ions within predefined m/z windows, ensuring comprehensive and unbiased peptide quantification and identification (Ludwig et al., 2018). Thus, DIA MS makes the identification of novel peptides possible and provides a deep understanding of protein abundance and post-translational modifications, which is profoundly important in clinical proteomics (Sajic et al., 2015).

In parallel, data-dependent acquisition (DDA) MS, accompanied by peptide labeling techniques such as tandem mass tags (TMT), has evolved as a versatile technique in clinical proteomics (Aljawad et al., 2023). TMT labeling allows simultaneous comparison of peptide abundances across multiple samples in a single MS run, providing the highest accuracy of all relative quantitative proteomic techniques. However, it comes with high costs and strict experiment design requirements (Brenes et al., 2019).

DIA MS is usually performed without peptide labeling, termed label-free quantification (LFQ). DIA-LFQ methods are cheaper and require fewer sample preparation steps, but the accurate quantification of low-abundance proteins is limited (Rozanova et al., 2021). Thus, both DIA-LFQ and DDA-TMT methods provide unique strengths and are recognized methods in clinical proteomics.

To maximize the potential of clinical proteomics, analyzing larger multi-center patient cohorts is necessary to increase statistical power and achieve more robust results, especially for identifying rare disease subtypes (Hernández et al., 2014). (Zhang et al., 2022). However, integrating patient-derived MS data distributed across multiple research institutions can be problematic due to privacy concerns. Proteomics data provide in-depth insights into an individual's protein abundance profile and post-translational modifications and, similar to transcriptomics data, can be subject to genotype reconstruction attacks (Geyer et al., 2021). This risk significantly increases when considering the possibility of detecting rare sequence variants (Fierro-Monti et al., 2022). Therefore, raw patient-derived MS data and proteomics profiles must be treated as confidential information.

Currently, the only way to collectively analyze distributed proteomic data without compromising patients' privacy due to direct data sharing and pooling is to combine individual study outcomes using meta-analysis techniques (Kaever et al., 2014).

Various methodologies are used, each presenting unique advantages and limitations. In general, meta-analysis performance improves with a larger sample size within studies and a higher number of studies (Turner et al., 2013) or with the availability of raw data for combined re-analysis (Adamowicz et al., 2023), which is challenging to achieve in proteomics. Common meta-analysis techniques include Fisher's method (Fisher, 1925; Kaever et al., 2014), Stouffer's method (Kaever et al., 2014; Stouffer et al., 1949), the random effects model (REM) (Choi et al., 2003; Haidich, 2010), and RankProd (Breitling et al., 2004; Hong et al., 2006).

A consistent limitation of most meta-analyses is the underlying assumptions about p-value or effect size distributions, which might not be realistic. Additionally, meta-analyses face challenges related to heterogeneity from variations in experimental design, sample characteristics, equipment used for peptide separation and MS data acquisition, and data preprocessing methods (Makinde et al., 2021). They cannot fully account for cohort differences, such as target class imbalance or variations of covariate distributions (Higgins and Thompson, 2002; Zolotareva et al., 2021). Differences in data processing steps, such as normalization, may also significantly impact the meta-analysis results (Bullard et al., 2010).

To enable privacy-preserving analysis of distributed proteomic data owned by multiple institutions while prioritizing data privacy and ensuring robust results despite data heterogeneity, we suggest applying federated learning (McMahan et al., 2017) in combination with privacy-enhancing technologies like secure multi-party computation (SMPC) or additive secret sharing (Cramer et al., 2015). Recently, the power of a hybrid approach based on federated learning and SMPC to protect privacy during data integration in transcriptomics has been demonstrated by *Flimma* (Zolotareva et al., 2021), a privacy-aware tool for differential gene expression analysis of decentralized data. However, *Flimma* cannot be applied to proteomics data due to its variance estimation approach from *limma voom* (Law et al., 2016), which is tailored for RNA-seq and unsuitable for MS data. Also, *Flimma* employs normalization approaches designed for count data, and it does not implement filtering procedures necessary for proteomics data, nor can it handle inputs with missing values. Missing values are intrinsic to proteomics data due to instrument sensitivity and method design (Lazar et al., 2016) or the stochastic sampling nature of MS, resulting in inconsistent detection of low-abundance proteins (Collins et al., 2017).

To fulfill an unmet need for a privacy-aware approach tailored specifically for MS-based proteomics, we designed FedProt — a federated learning-based tool for collaborative differential protein abundance analysis of distributed data. FedProt is based on *DEqMS*, a state-of-the-art *limma's* modification for estimating variance that enhances overall performance (Zhu et al., 2020). Unlike *DEqMS* and other tools requiring data centralization, FedProt, by design, preserves the patients' privacy since the protein abundance profiles always remain in the local environments of the collaboration parties and are never shared externally.

To evaluate FedProt, we employed two most commonly used approaches, LFQ and TMT, and created two multi-center datasets: an LFQ bacterial dataset from 5 independent centers and a TMT human serum dataset from 3. We also used simulated data to test FedProt’s behavior under data imbalance. Our results demonstrate that regardless of data imbalance or batch effects, FedProt always delivers exactly the same results as the original *DEqMS* workflow, a great advantage over other meta-analysis methods.

# Results

## FedProt overview

FedProt represents the mathematical equivalent of *DEqMS* (Zhu et al., 2020), the accurate variance estimation workflow for mass spectrometry-based proteomics data. To protect the privacy of patient-derived data, FedProt utilizes the hybrid approach of federated learning (McMahan et al., 2017) and additive secret sharing (Cramer et al., 2015), similar to Flimma (Zolotareva et al., 2021). The FedProt workflow overview is shown in Figure 1.

Federated learning is a machine learning paradigm that increases data privacy by allowing multiple parties to collaboratively train a model without revealing their sensitive data to each other (McMahan et al., 2017). That is achieved by splitting the computational workflow into steps each participant performs on their local data. The participants (clients) exchange only model parameters through a central trusted server orchestrating the computations. To implement the federated learning scheme, all participants run the same application instance, accessing only its local data and communicating local results with the server. The central server (Coordinator) collects local results from clients, aggregates them into global results, and returns them to clients for a new step. The key point of federated learning is that the intermediate local or global results are constructed to minimize the risk of sensitive local data reconstruction.

To further enhance privacy and protect local statistics and parameters from being revealed to the server, we use additive secret sharing (Cramer et al., 2015). In this method, each client generates multiple masks and communicates masks and masked data to the other parties, ensuring that no single party receives more than one piece of the data from each of the other clients (green arrows in Figure 1). Each piece is encrypted with the recipient's public key, all received data summed by the receiving parties, and sent to the Coordinator (blue arrows in Figure 1). The Coordinator receives summarized data and computes and broadcasts the global model to all clients (black arrows in Figure 1). This scheme allows global aggregation of local results without revealing any local values,enhancing privacy compared to a pure federated learning scheme (see Methods for further details).

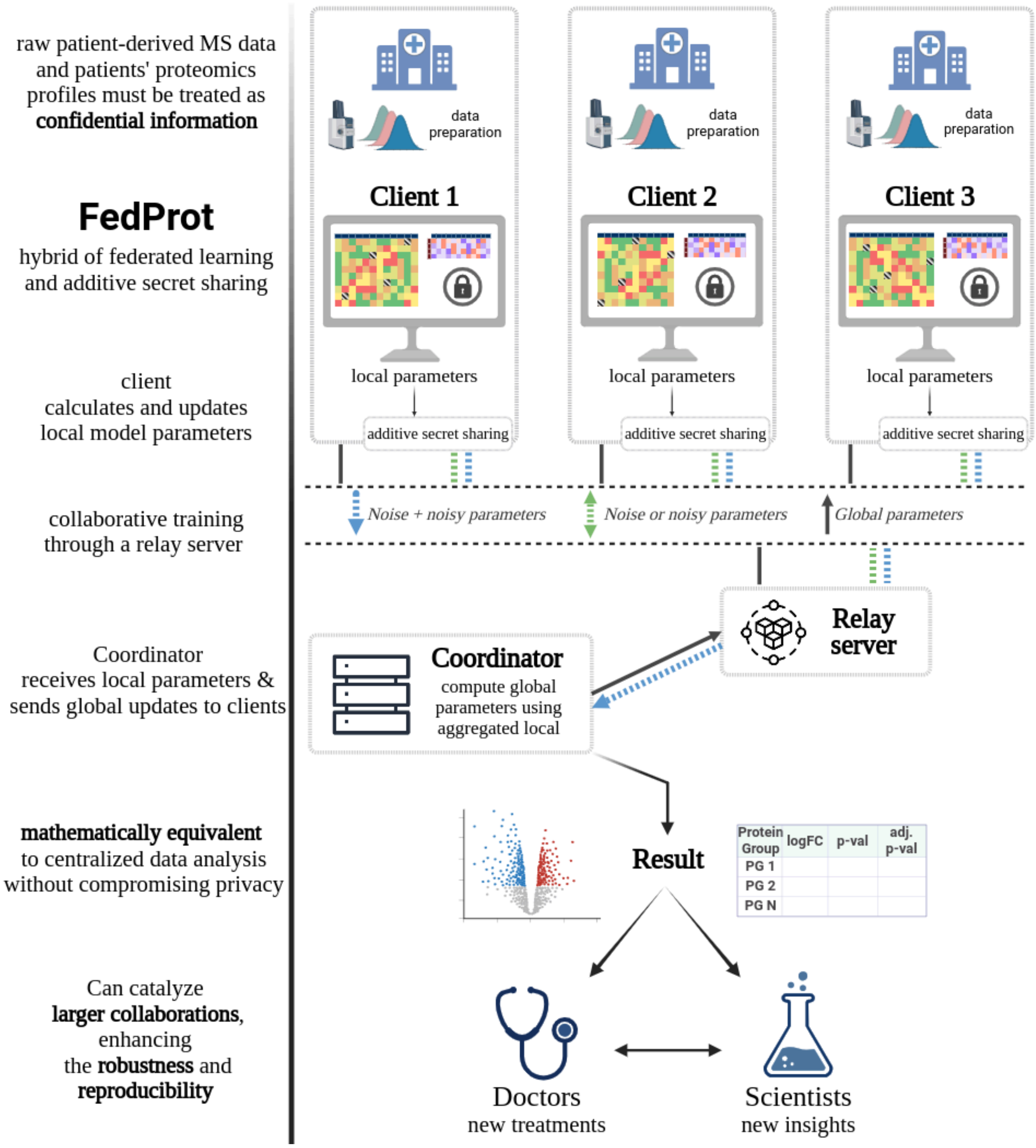

**Figure 1. FedProt Workflow Overview.**

**Data Preparation:** Data owners collect and preprocess MS data, obtain protein intensity and peptide count matrices, and define design matrices before participating as clients.

**Federated Learning:** Clients communicate with the central server (*Coordinator*) to collaboratively train a global model without revealing their individual datasets, but through the exchange of local model parameters via the Relay server. "Model" here refers to the set of intermediate parameters generated during computation. The clients protect their local parameters using additive secret sharing, where encrypted parts of masked data are exchanged among clients (green arrows) through the Relay server, then these parts are added together and relayed again through the Relay server to the Coordinator (blue arrows) to update the global model without exposing individual local datasets. The Coordinator returns updated global parameters to clients (black arrows).

**Result:** After all federated computations, all clients receive the results mathematically equivalent to the results of centralized analysis of pooled dataset with *DEqMS* formatted as a table with abundance fold-changes, confidence intervals, and adjusted p-values.

To make this decentralized approach and its complex infrastructure available to a broad community, we implemented FedProt as a web-based app with a user-friendly graphical interface. FedProt is published as a certified app in the FeatureCloud (Matschinske et al., 2023) app store with documentation and quick-start guidelines. To initialize the workflow analysis, the coordinator sets up a workflow and invites collaborating parties to join using an invitation token. All parties should be registered in featurecloud.ai, download and run the FedProt app. The coordinator is responsible for setting the analysis parameters (e.g., expected number of participants, filtering parameters) and sharing the analysis with them. Each participant should specify paths to local input data, including three .tsv files containing (i) patients' protein intensity profiles, (ii) design matrices featuring class labels and covariates, and (iii) matrices of minimal peptide count across all samples for each unique protein group.

The FedProt federated workflow starts when all clients, invited by the coordinator, join and provide correctly formatted inputs (see Methods). Upon its successful completion, each client receives a table with expression fold-changes, confidence intervals, and adjusted p-values, in the same format as the *DEqMS* output. The FedProt approach allows us to obtain the same result as centralized pooled data analysis while implementing strong privacy-preserving measures, ensuring no patient-level data is shared and exchanged parameters are masked.

## Evaluation approach

Due to privacy regulations, finding publicly available multicenter real patient-derived data suitable for evaluating FedProt was challenging. Therefore, specifically for this benchmark, we created two real-world test datasets, one quantified using LFQ and the other using TMT (Table 1).

The LFQ-based dataset included 118 E. coli colonies cultured under two growth conditions, simulating case and control groups. Of these samples, 98 were unique and uniformly distributed between five independent labs, and four samples were measured by all labs for quality control purposes (Supplementary figure S1).

The TMT-based dataset consisted of three cohorts, each including ten serum samples from individuals with focal segmental glomerulosclerosis (FSGS) and ten control samples. Within each cohort, the samples were randomly distributed between two TMT batches, with five samples from each group.

To evaluate FedProt, each center's raw mass spectra were separately quantified using MaxQuant (for the TMT dataset) (Tyanova et al., 2016) and DIA-NN (for the LFQ dataset) (Demichev et al., 2020) software with the same settings and FASTA files as a database. We assumed that collaborating parties could agree on using a uniform data preprocessing protocol.

Following quantification, pooled data were centrally analyzed using the *DEqMS* method (Zhu et al., 2020) to establish a baseline (ground truth). We then compared this ground truth against the results of FedProt and also against the results derived from four meta-analysis methods: Fisher's (Fisher, 1925) and Stouffer's methods (Stouffer et al., 1949), the random effects model (REM) (Choi et al., 2003), and RankProd (Breitling et al., 2004; Hong et al., 2006).

**Table 1.** Characteristics of datasets for FedProt evaluation. Number of samples in each cohort in each condition. A — bacterial dataset, B — human serum dataset. FSGS — focal segmental glomerulosclerosis group. M9 is the medium used to grow *E. coli*.

| **A. Bacterial dataset** | | |
|---|---|---|
| | **M9 Pyruvate** | **M9 Glucose** |
| Lab A | 10 | 10 |
| Lab B | 10 | 9 |
| Lab C | 9 | 10 |
| Lab D | 10 | 10 |
| Lab E | 10 | 10 |
| **B. Human serum dataset** | | |
| | **Control** | **FSGS** |
| Center 1 | 10 | 10 |
| Center 2 | 10 | 10 |
| Center 3 | 10 | 10 |

When analyzing proteomics data from different sources, we encounter incomplete overlap of quantified protein groups between cohorts (see Supplementary figure S2). The centralized *DEqMS* method allows for analyzing almost all features in the aggregated dataset, except for those that do not pass the filter on the number of available measurements for each target class. Because of privacy concerns, FedProt poses an additional restriction to analyzed features and excludes protein groups with only one measurement per cohort that can be used for a reconstruction attack (Zolotareva et al., 2021).

Meta-analyses can analyze different numbers of features. For example, the Stouffer and RankProd methods we used can analyze only protein groups present in all cohorts. The Fisher method requires the protein group to be present in at least two cohorts. The REM can use all input protein groups, although, as we will observe in the later analysis, this does not improve the quality of its results.

Supplementary figure S3 illustrates this limitation and quantifies the features lost due to decentralization for the bacterial and human serum datasets. After the differential abundance analysis, we computed evaluation metrics using features that the methods with the fewest analyzable features could process; in our case, these were RankProd and Stouffer's methods.

Additionally, to investigate the effect of data imbalance on the results of decentralized methods, we created simulated data with increasing levels of imbalance across cohorts (Methods).

## Deviations in the results of decentralized methods

FedProt produced results that matched the results of the centralized *DEqMS* workflow in all tests for both datasets. This was evident in the consistency between the mean absolute difference for adjusted p-values and log-fold-change values, as the maximum absolute differences were negligible (no greater than $4*10^{-12}$, see Table 2, Supplementary table S1).

In contrast, the log-transformed p-values of the meta-analysis methods demonstrated notable deviations from the ground truth, with mean differences ranging from 3.44 to 15.72 for the LFQ dataset and 0.50 to 1.1 for the TMT dataset (Table 2, Figure 2). Similarly, the log-fold-change results from the meta-analyses demonstrated larger differences with maximal absolute differences up to 0.2 (Supplementary table S1 and figure S4). To compare log-fold changes, we only used the results of REM and Fisher's method, since Fisher's method calculates it using the same "mean" approach as Stouffer's method and RankProd.

**Table 2.** The mean and maximum absolute differences between the negative log-transformed BH-adjusted p-values of centralized *DEqMS* and FedProt or tested meta-analysis approaches. The lowest differences are shown in bold font.

| Dataset | **Method** | **Mean difference** | **Maximal difference** |
|---|---|---|---|
| **Bacterial dataset** | **FedProt** | **4.43E-13** | **3.46E-12** |
| | **Fisher** | 4.00 | 26.78 |
| | **Stouffer** | 3.44 | 24.99 |
| | **REM** | 15.72 | 262.62 |
| | **RankProd** | 14.25 | 83.68 |
| **Human serum dataset** | **FedProt** | **1.36E-13** | **6.59E-13** |
| | **Fisher** | 0.50 | 2.79 |
| | **Stouffer** | 0.57 | 2.59 |
| | **REM** | 0.59 | 11.64 |
| | **RankProd** | 1.07 | 8.98 |

This superior consistency of FedProt and centralized *DEqMS* results proves that the federated approach can achieve the same results as the centralized model but with the significant advantage of privacy protection.

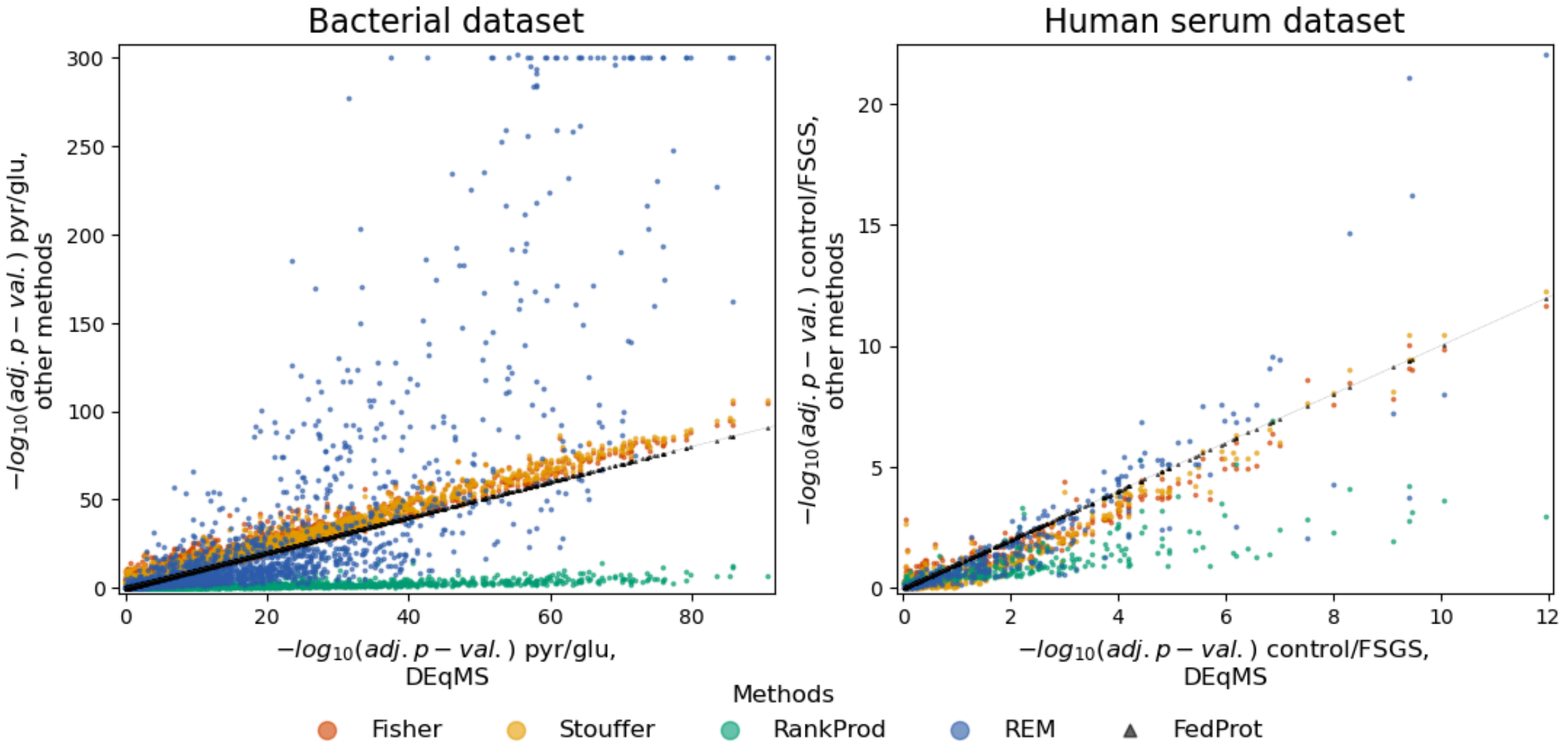

**Figure** ***2.***
The comparison of negative log-transformed adjusted p-values computed by FedProt or meta-analysis methods (y-axis) with the centralized *DEqMS* analysis (x-axis). -log10(adj.p-val) is negative log-transformed count-adjusted BH-method adjusted p-values. The thin black line is the diagonal.

## The consistency of differentially abundant protein lists

Averaged absolute differences quantify the discrepancy between the centralized and decentralized methods for all protein groups, given that the errors computed for both differentially and non-differentially abundant proteins are treated equally during statistical analysis (Tables 2, S1). However, in many studies, the exact p-values and effect sizes are not as crucial as accurate identification and consistent ranking of differentially abundant proteins. Therefore, in addition to overall consistency assessments, we compared the lists of the most strongly and significantly differentially abundant proteins detected by decentralized methods.

As before, the lists of differentially abundant proteins identified by decentralized methods were compared to the ground truth list obtained by the centralized *DEqMS* method. We applied $|\log 2\mathrm{FC}|$ > 0.5 and adjusted p-value of 0.05 thresholds for the bacterial dataset, and for the human serum dataset — 0.25 and 0.05, respectively. The $|\log 2\mathrm{FC}|$ thresholds were selected based on their distributions in the datasets. Method performances in terms of false positives (FP), false negatives (FN), and Jaccard similarities are presented in **Table 3**.

Of all tested methods, only FedProt identified exactly the same differentially abundant proteins as centralized *DEqMS*. Regardless of the method used, the outputs of all meta-analysis methods always contained FP and FN. And despite the smaller number of protein groups in the human serum data, the deviation on this dataset for meta-analyses from ground truth was higher.

**Table 3**. Jaccard similarity coefficients, the number of false positives (FP), and the number of false negatives (FN) obtained on bacterial and human serum datasets for FedProt and meta-analysis methods.

| Dataset | Method | FP | FN | Jaccard similarity coefficient |
|---|---|---|---|---|
| **Bacterial dataset** $\|\log2\mathrm{FC}\|$ > 0.5, adj.p-value < 0.05 | **FedProt** | **0** | **0** | **1** |
| | **Fisher** | 3 | 4 | 0.99 |
| | **Stouffer** | 3 | 4 | 0.99 |
| | **REM** | 8 | 15 | 0.96 |
| | **RankProd** | 1 | 111 | 0.81 |
| **Human serum dataset** $\|\log2\mathrm{FC}\|$ > 0.25, adj.p-value < 0.05 | **FedProt** | **0** | **0** | **1** |
| | **Fisher** | 2 | 6 | 0.92 |
| | **Stouffer** | 2 | 13 | 0.85 |
| | **REM** | 1 | 13 | 0.86 |
| | **RankProd** | 1 | 33 | 0.67 |

Additionally, by ranking proteins based on the adjusted p-values, we assessed how method performance shifted with changing the number of top differentially abundant proteins (Figure 3). We did that because identifying a limited number of significantly differentially abundant proteins is often a task in studies like biomarker discovery. Like in previous tests, FedProt consistently matched the results of the centralized approach in both test scenarios, outperforming all meta-analysis methods.

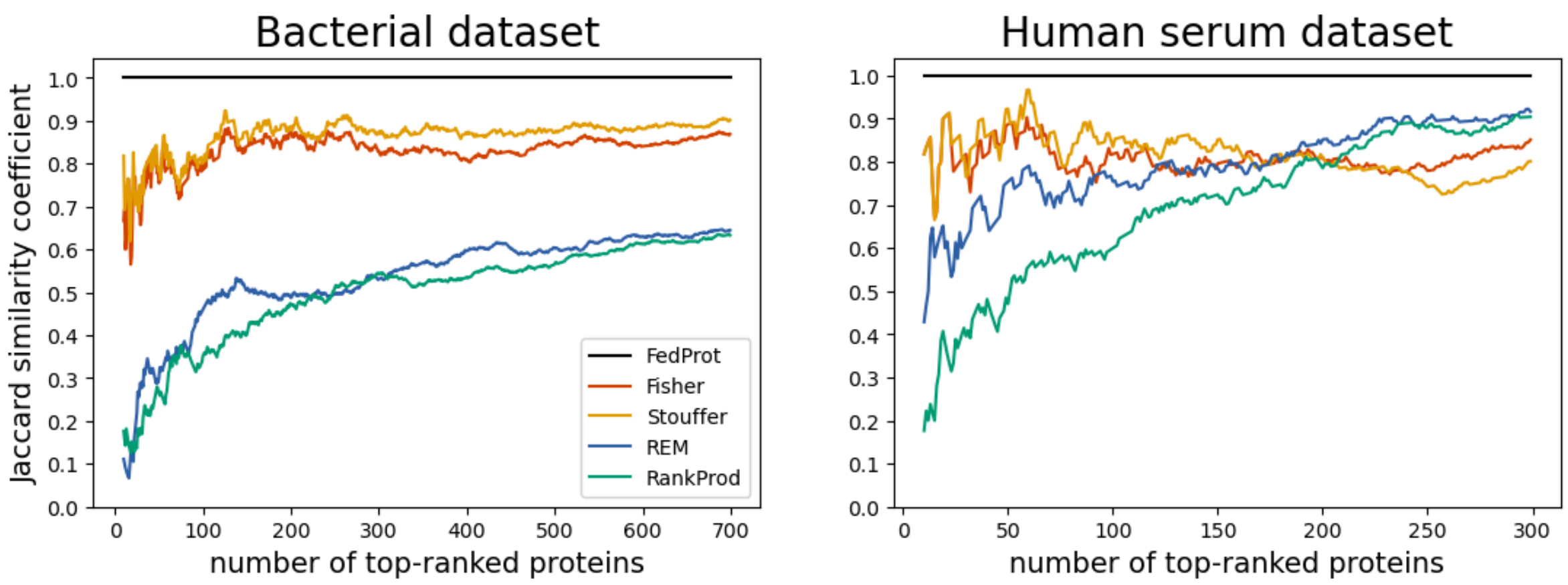

**Figure 3.**
The dependency of the Jaccard similarity coefficient on the number of top-ranked proteins identified by the centralized DEqMS and decentralized approaches. Proteins were ranked based on their decreasing negative log-transformed BH-adjusted p-values and not filtered by log2FC.

## Robustness against data imbalance

We further tested FedProt's results reliability even when faced with the challenge of data imbalance using simulated data. We generated three protein intensity matrices with two conditions (groups A and B) for balanced, mild imbalanced, and strong imbalance scenarios (Table 4) for 6000 proteins and 600 samples using the same approach described in (Wang et al., 2021). Batch effects were introduced using the ComBat model (Johnson et al., 2007), and missing values were added as per Jin et al., 2021 (Jin et al., 2021). Additionally, we simulated a confounder in condition B, varying its frequency across cohorts in three scenarios. Since our focus was to investigate the effect of data imbalance on the results, counts were not simulated, and the analyses were concluded without count adjustment. Each simulation and analysis was repeated 50 times.

**Table 4.** Characteristics of the simulated datasets used to evaluate the effect of data imbalance. Number of samples in each cohort in each condition, for confounder column — proportion of samples among condition B samples.

| | Cohorts | | | Condition A | | | Condition B | | | in B — frequency of samples with the confounder | | |
|---|---|---|---|---|---|---|---|---|---|---|---|---|
| | B 1 | B 2 | B 3 | B 1 | B 2 | B 3 | B 1 | B 2 | B 3 | B 1 | B 2 | B 3 |
| **Balanced** | 200 | 200 | 200 | 100 | 100 | 100 | 100 | 100 | 100 | 0.6 | 0.6 | 0.6 |
| **Mild imbalanced** | 90 | 140 | 370 | 36 | 91 | 185 | 54 | 49 | 185 | 0.4 | 0.5 | 0.66 |
| **Strong imbalanced** | 40 | 80 | 480 | 32 | 28 | 288 | 8 | 52 | 192 | 0.2 | 0.5 | 0.7 |

Regardless of data imbalance, FedProt produced results that closely matched those from the centralized DEqMS workflow in all scenarios. As in previous tests, the mean absolute differences for adjusted p-values and log-fold-change values were exceptionally small, with the maximum absolute difference not exceeding $6.7*10^{-13}$ (see Tables 5, S2).

As the degree of data imbalance increases, the maximum and mean absolute differences for meta-analysis methods consistently increase (see Tables 5, S2, Figure 4). The strong imbalance significantly affects meta-analyses results not only in adjusted p-values but also in log-fold changes (Table S2, Figure S5). In contrast, FedProt stably demonstrated the absence of errors and achieved robust performance.

**Table 5.** The mean and maximum absolute differences between the negative log-transformed adjusted p-values of centralized *DEqMS* and FedProt or selected meta-analysis approaches. The lowest differences are shown in bold font. The generation of simulated data and the subsequent data analysis were repeated 50 times — mean and standard deviation for the mean absolute differences for these analyses results are provided.

| Dataset | Method | Mean difference | Maximal difference |
|---|---|---|---|
| **Simulated, balanced** | **FedProt** | **3.22E-15 ± 3.52E-16** | **1.85E-13 ± 8.71E-14** |
| | **Fisher** | 0.12 ± 0.01 | 12.50 ± 4.14 |
| | **Stouffer** | 0.14 ± 0.01 | 8.99 ± 3.39 |
| | **REM** | 0.15 ± 0.01 | 18.90 ± 4.79 |
| | **RankProd** | 0.78 ± 0.02 | 27.60 ± 4.00 |
| **Simulated, mild imbalance** | **FedProt** | **6.00E-15 ± 6.22E-16** | **2.67E-13 ± 7.30E-14** |
| | **Fisher** | 0.14 ± 0.01 | 13.70 ± 4.65 |
| | **Stouffer** | 0.20 ± 0.01 | 8.70 ± 2.18 |
| | **REM** | 0.19 ± 0.01 | 23.50 ± 4.06 |
| | **RankProd** | 0.84 ± 0.02 | 32.10 ± 3.93 |
| **Simulated, strong imbalance** | **FedProt** | **1.33E-14 ± 1.69E-15** | **6.63E-13 ± 2.74E-13** |
| | **Fisher** | 0.15 ± 0.01 | 13.20 ± 4.67 |
| | **Stouffer** | 0.41 ± 0.06 | 29.20 ± 8.11 |
| | **REM** | 0.22 ± 0.01 | 26.20 ± 4.94 |
| | **RankProd** | 0.91 ± 0.02 | 34.40 ± 5.62 |

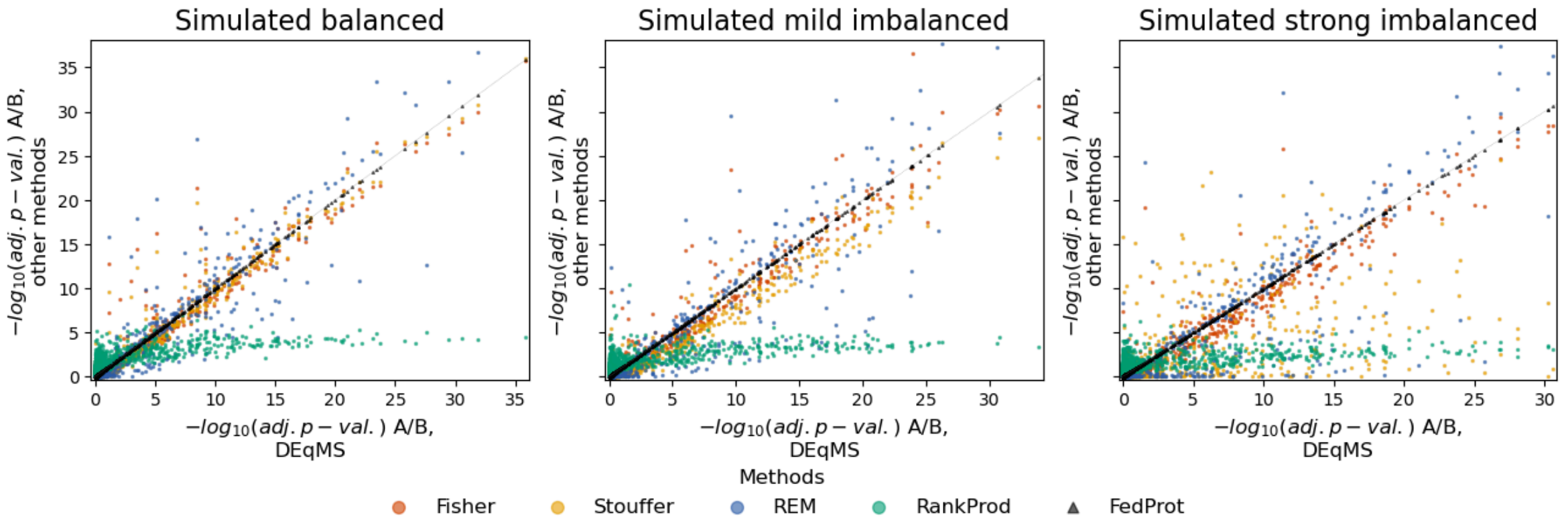

**Figure 4.**
The comparison of negative log-transformed adjusted *DEqMS* p-values computed by FedProt or meta-analysis methods (y-axis) with centralized analysis (x-axis) for one out of 50 analysis runs for each scenario.

As before, we analyzed how the data imbalance affects the list of differentially abundant proteins. The protein groups with |log2FC| > 1 and passing an adjusted p-value threshold of 0.05 were considered differentially abundant.

Regardless of the method used, the outputs of all meta-analysis methods always contained FP and FN, and their counts grew with an increasing degree of imbalance in data (Table 6). Importantly, data imbalance does not impact FedProt's results. The results remain stable under varying degrees of imbalance.

**Table 6.** Jaccard similarity coefficient, the number of false positives (FP), and the number of false negatives (FN) obtained for FedProt and meta-analysis methods. For the simulated datasets the mean values and standard deviation for 50 runs are reported. The values corresponding to the best performance between all methods are highlighted in bold font.

| Dataset | Method | FP | FN | Jaccard similarity coefficient |
|---|---|---|---|---|
| **Simulated balanced** $\lvert \log2FC \rvert > 1$, adj.p-value < 0.05 | **FedProt** | **0.0 ± 0** | **0.0 ± 0** | **1.00 ± 0.00** |
| | **Fisher** | 2.3 ± 1.5 | 2.2 ± 1.4 | 0.95 ± 0.02 |
| | **Stouffer** | 2.3 ± 1.5 | 2.6 ± 1.4 | 0.94 ± 0.02 |
| | **REM** | 6.2 ± 2.7 | 7.4 ± 2.5 | 0.84 ± 0.04 |
| | **RankProd** | 11.1 ± 3.4 | 1.0 ± 0.9 | 0.87 ± 0.03 |
| **Simulated mild imbalanced** $\lvert \log2FC \rvert > 1$, adj.p-value < 0.05 | **FedProt** | **0.0 ± 0** | **0.0 ± 0** | **1.00 ± 0.00** |
| | **Fisher** | 18.2 ± 3.9 | 17.1 ± 4.6 | 0.64 ± 0.04 |
| | **Stouffer** | 17.8 ± 3.9 | 17.6 ± 4.5 | 0.64 ± 0.04 |
| | **REM** | 8.1 ± 2.6 | 10.3 ± 3.8 | 0.79 ± 0.04 |
| | **RankProd** | 38.9 ± 6.7 | 16.0 ± 4.3 | 0.54 ± 0.04 |
| **Simulated strong imbalanced** $\lvert \log2FC \rvert > 1$, adj.p-value < 0.05 | **FedProt** | **0.0 ± 0** | **0.0 ± 0** | **1.00 ± 0.00** |
| | **Fisher** | 40.0 ± 5.9 | 29.4 ± 4.7 | 0.45 ± 0.04 |
| | **Stouffer** | 33.4 ± 4.1 | 36.3 ± 5.6 | 0.41 ± 0.05 |
| | **REM** | 7.8 ± 2.8 | 15.3 ± 3.3 | 0.75 ± 0.04 |
| | **RankProd** | 147.0 ± 11.8 | 28.2 ± 4.8 | 0.25 ± 0.02 |

Regarding the dependence of the Jaccard similarity index on the number of top-ranked protein groups, again, meta-analyses showed higher discrepancy at the top of the list and it grows with the increase of data imbalance (Figure 5). At the same time, FedProt's results completely match the results of the central analysis, regardless of dataset or imbalance.

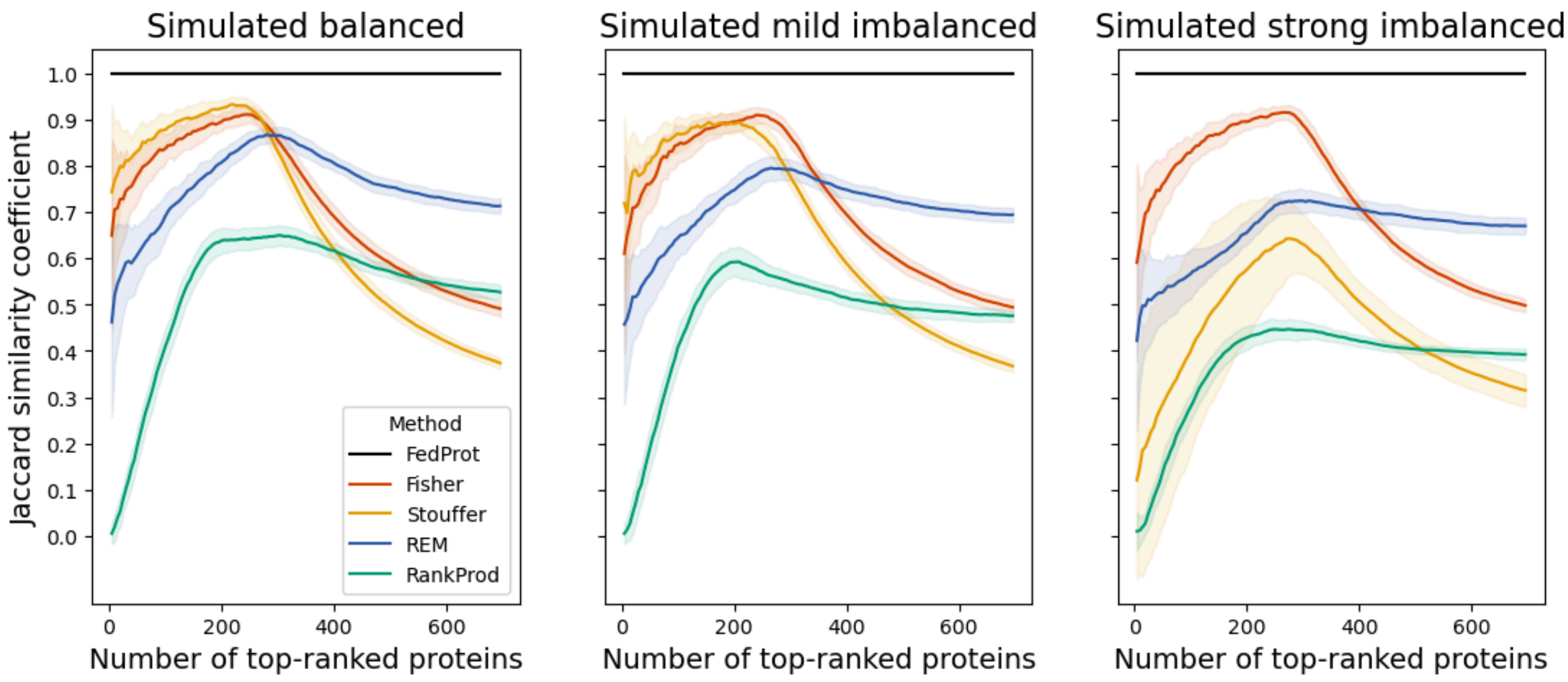

**Figure 5.**
The dependency of the Jaccard similarity coefficient on the number of top-ranked proteins. Proteins were ranked based on their decreasing negative log-transformed BH-adjusted p-values. The generation of simulated data and the subsequent data analysis were repeated 50 times — aggregated results reported.

## Handling of batch effects

Batch effects are a significant issue in data analysis since they can mask true biological differences and lead to incorrect conclusions. It is crucial to handle batch effects properly to ensure the validity of differential abundance analysis results. For both bacterial and human serum datasets, principal component analysis revealed large differences between samples from different cohorts, exceeding differences within the same cohort (Figure S6). This lab-specific batch effect was particularly noticeable when comparing samples from labs A and B, who performed cell lysis, and those from labs C, D, and E, who worked with centrally prepared cell lysates.

Since unaccounted lab-specific batch effects can severely confound the analysis results (Čuklina et al., 2021), researchers either adjust the data to remove batch effects before the analysis or modify the model to account for them. *ComBat* (Johnson et al., 2007), a popular batch effect correction tool, does not accept data with missing values, making it unsuitable for multicenter proteomic data, where the number of rows with missing values increases with the number of participating laboratories. As imputation reliability is questionable, the recent HarmonizR workflow has been developed to handle inputs with missing values (Voß et al., 2022). However, the current version of HarmonizR cannot account for confounders, which is necessary to avoid overcorrection in imbalanced data.

To ensure FedProt's approach to handling differential expression analysis and batch effect adjustment is not inferior to popular approaches involving direct data adjustment, we compared its results with those of DEqMS applied to pooled data after batch effect correction using the *removeBatchEffect* function from the limma R package (Ritchie et al., 2015), which is able to account for covariates and correct the data with missing values. Similar to Flimma (Zolotareva et al., 2021) and the limma R package, to account for batch effects in the analysis involving $m$ clients, it selects one client that serves as the reference batch and adds to the design matrix $m-1$ binary covariates modeling expression changes of each client w.r.t. the reference batch.

Including batch factors in the design is recommended to correctly assess standard errors (Ritchie et al., 2015). Using pre-corrected data as “batch effect free” in the analysis may lead to exaggerated confidence (Nygaard et al., 2016). We observed these effects on adjusted p-value results by comparing two DEqMS analyses on centrally aggregated data: one corrected using limma’s *removeBatchEffect* function and the other on uncorrected data with batch factors included in the model (Figure S7).

Nevertheless, we can compare log-fold changes to evaluate FedProt against central analysis on batch effect corrected data and thus visualize removing the batch effect. For both datasets, log fold-changes were perfectly correlated (r=1, rho=1) between FedProt on non-corrected data and centralized *DEqMS* workflow on corrected data with maximal absolute differences no greater than 5.5E-14 (Figure S8).

# Discussion and conclusion

In this study, we introduce FedProt — the first privacy-preserving tool for federated differential protein abundance analysis. FedProt is based on *DEqMS* for the accurate variance estimation in MS data and utilizes a hybrid approach of federated learning and additive secret sharing (Cramer et al., 2015; Zolotareva et al., 2021) to ensure patient-derived data privacy. Building on this foundation, FedProt is poised to revolutionize the field through enabling larger distributed proteomics data analysis to a broader audience.

FedProt is a user-friendly tool that increases data analysis robustness while keeping the data at hospitals' legally safe harbors, thus significantly minimizing privacy risks without compromising accuracy. FedProt increases sample sizes, enhancing the statistical power of differential abundance analyses, and effectively managing challenges like batch effects and missing data.

To simulate a realistic multi-center study and evaluate FedProt, we created the bacterial LFQ dataset (PXD053812) and the human serum TMT dataset (PXD053560). The LFQ dataset consists of 118 *Escherichia coli* samples (2 conditions, five cohorts), and the TMT human serum dataset of 60 samples (2 conditions, three cohorts). Each cohort (research center) measured samples using the available MS analysis equipment. With this, we have filled a significant gap by offering datasets that allow for multiple cohort modeling, ensuring manageable batch effects and thereby facilitating FedProt performance evaluation.

Our evaluation confirmed that FedProt's results are equivalent to those from the original *DEqMS* method when applied to centralized and pooled data. In all tests, particularly in identifying top differentially abundant proteins, FedProt’s results consistently matched the centralized approach results, surpassing all tested meta-analysis methods. Additionally, FedProt demonstrated resilience to a sample size imbalance between cohorts, the ability to work with data with missing values, and accounting for batch effects.

However, our study is not devoid of limitations. Firstly, the current version of FedProt only supports LFQ and TMT proteomics data and two normalization methods. However, FedProt's design is inherently adaptable, and the current workflow can be concluded without count adjustment. It allows future extensions to other data types like phosphoproteomics (Xu et al., 2019) or metabolomics (Myint et al., 2017), and to multi-omics data analysis (Lehmann et al., 2021).

Secondly, the federated learning approach enables privacy-preserving analysis of distributed data but cannot guarantee absolute privacy alone. By using additive secret sharing (Cramer et al., 2015; Matschinske et al., 2023), FedProt enhances privacy protection compared to pure federated learning, ensuring the local parameters' original values remain hidden from the central server. Additionally, it includes built-in checks and alerts for client-side data anomalies, which will stop the computation if the number of clients involved exceeds the total number of samples. Besides, such scenarios are very unlikely and the likelihood of reconstruction attacks is extremely low (Melis et al., 2019; Nasirigerdeh et al., 2021).

Overall, FedProt provides enhanced privacy protection compared to traditional centralized analysis at the cost of negligibly small errors compared to errors of meta-analyses. It is a promising approach with the potential to facilitate larger-scale, privacy-preserving multi-center collaborations in clinical proteomics.

# Methods

## FedProt workflow

FedProt is based on the accurate variance estimation workflow of *DEqMS* (Zhu et al., 2020) for MS data and employs the hybrid approach of federated learning and additive secret sharing (Cramer et al., 2015) in a manner similar to *Flimma* (Zolotareva et al., 2021). Federated learning is a machine learning paradigm in which models are trained on multiple devices (clients) without centralizing data, enabling collaborative learning with increased data privacy. Clients securely store and analyze their local data, and exchange intermediate results with the trusted server (Coordinator) that aggregates local results to global.

Before the federated analysis, each participant preprocesses its local dataset independently and is responsible for ensuring the quality and consistency of the data provided to the client app. We assume that all the participants used the same protocol to quantify and preprocess their local data.

The FedProt's federated analysis workflow is divided into six (or seven if normalization is needed) steps, of which four (or five) involve federated computations and require one or several rounds of communication between client and server (Figure 6).

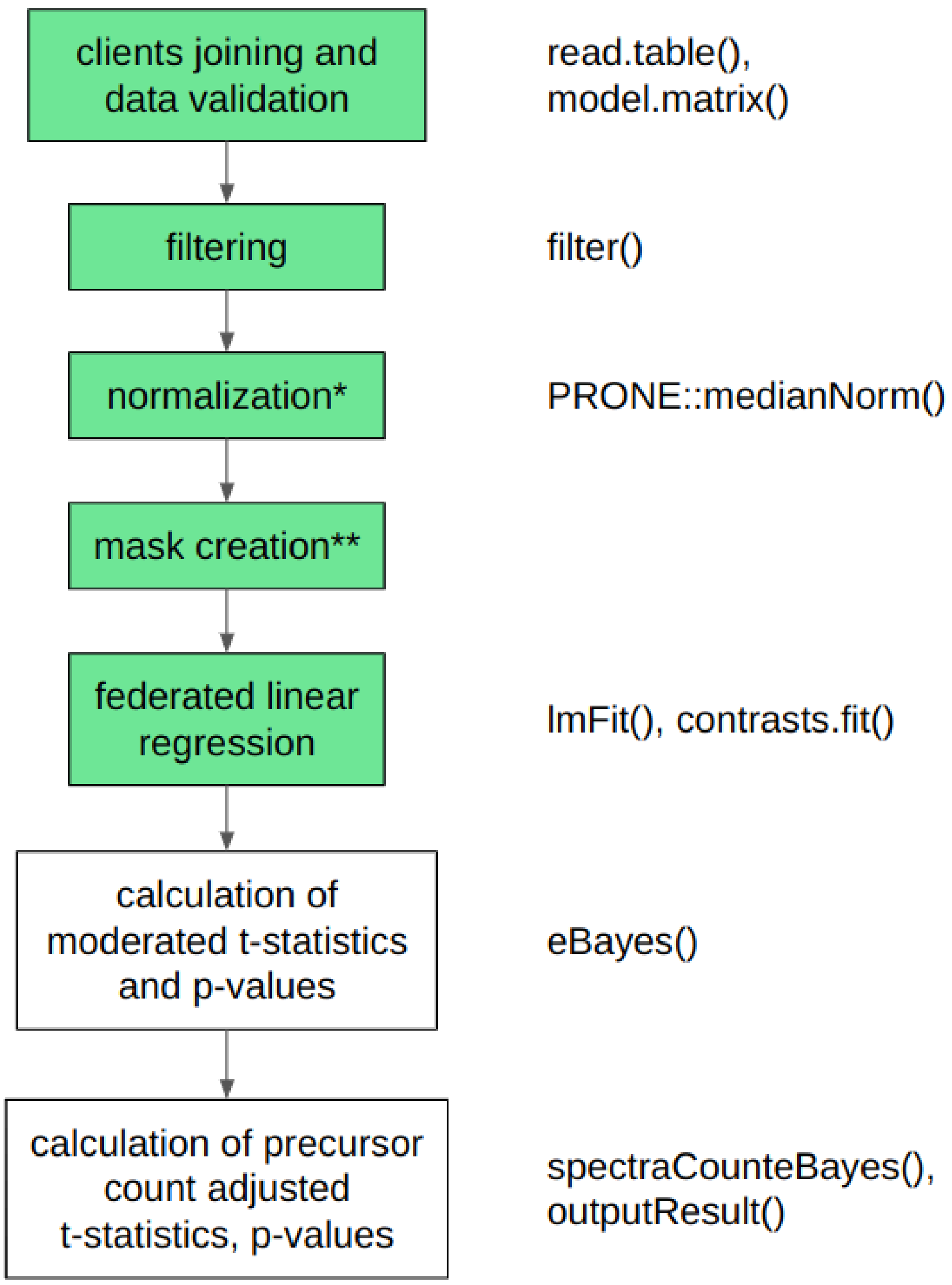

**Figure 6.**
Scheme of the FedProt workflow.
Steps that involve federated computations are shown in green. The corresponding stages of *DEqMS* workflow are shown on the right. Median normalization from the PRONE R package (https://github.com/lisiarend/PRONE) was used.

* — The normalization step is optional and could be turned off by the Coordinator. Because the data derived from a DIA LFQ experiment and processed through MaxLFQ protein quantification (Cox et al., 2014) are usually already normalized, so no additional preprocessing is needed except for filtering by protein group Q-value. In case of TMT data, also no additional preparation is needed, except for filtering out decoys, contaminant and reverse protein groups. The normalization by median across all centers and internal reference scaling inside each center can be performed during FedProt run. We suggest enabling the "match between runs" option during raw MS data quantification.
** — The step is for the federated approach only.

In the first step, $k > 2$ FedProt clients join the analysis initialized by the coordinator and validate their data. From each client $c_i$, where $i = 1, ..., k$, FedProt requires to have two tab-separated (.tsv) files:

1) A protein intensities matrix $Y^i_{\text{raw}}$ containing the intensities of $n^i$ proteins or protein groups (rows) detected in $m^i$ samples (columns). Protein groups detected only in a single sample in the cohort are replaced with NA (missing value), to avoid exchanging individual-level data in the next steps and protect data privacy.

2) A design matrix $X^i$ specifying which columns of the intensities matrix belong to which experimental conditions or groups. In the design matrix, each experimental condition or group should be coded as a binary variable, with 1 if the sample belongs to the group and 0 if not. For example, if we compare conditions A and B, the design should contain both A and B columns with 0 or 1 for each sample. Optionally, the design matrix can contain columns representing the covariates. The names of target class columns and covariates in local design matrices must match the names specified by the coordinator during the initialization of the study.

3) An optional file — a matrix with the number of quantified precursor peptides for each protein group $Pr^i_{min}$. For each client one value per protein group is required. For example, it can be found in the Precursor.Ids column in the DIA-NN report, or in the Peptide.IDs column in the MaxQuant report. If the clients do not have $Pr_{min}$, the computation will be completed after the sixth step, with no precursor peptide count adjustment.

Clients join the server and during the first step FedProt ensures that all the clients provide all necessary inputs. Each client sends to the server information about the number of samples $m^i$, the peptide-to-protein minimal counts $Pr^i_{min}$ and the list of protein groups $P^i$ they have. The server uses each client's $Pr^i_{min}$ matrices to get the global minimal number of quantified precursor peptides across all samples from all clients for each protein group, $Pr_{min}$; this is needed for the last step.

For each client $c^i$, the server updates the set of all detected protein groups $P$ across all clients participating in the analysis. After that clients receive from the server the set of all protein groups $P$, and the list of variables $v$ accounted for in the model (target classes, list of cohorts, and, optionally, covariates). If the protein group is not in the client data, it is created and filled with NAs in the $Y^i_{raw}$.

The list of variables $v$ is used to update the design matrix $X^i$ and include columns representing batches to account for batch effects. Each client updates the design matrix $X^i$ and adds cohort effects to it based on the list of variables $v$. The cohort effects are added as binary columns, where the first client $c^1$ represents the reference batch, as in *limma*, and the corresponding column is not included in the design matrix. If the coordinator participates in the computation as a client, it becomes the first client $c^1$.

In the second step, FedProt applies filters. One of them is to filter out protein groups with too many missing values. Each client $c^i$ calculates the number of samples with missing values per target class for each protein group and shares it with the server. The server computes global fractions of missing values per protein group and target class and the protein groups not detected in more than $f$ fraction of each target class samples in both target classes are removed from $P$. The value of $f$ is set to 0.8 by default and can be adjusted by the coordinator. The next optional filter removes protein groups supported only by 1 peptide precursor, this filter can be enabled by the coordinator.

The next step is the normalization step, this step is optional and depends on the coordinator settings. Currently, two types of normalization are implemented. The first one is median normalization. For that each client calculates the median intensity across all protein groups for each sample $j$ in their dataset, $Med_j^i$. Client's median average $\overline{Med}^i$ is sent to the coordinator. The coordinator then calculates global weighted mean of client's sample medians:

$$\overline{Med} = \frac{\sum_{i=1}^{k} \overline{Med}^i \cdot m^i}{\sum_{i=1}^{k} m^i}$$

Once the global median mean is computed, the coordinator broadcasts it to clients. Each client's j-th sample intensity values are adjusted based on this value:

$$Y^i_{norm,j} = \frac{Y^i_{raw,j}}{Med^i_j} \cdot \overline{Med}.$$

The second normalization is internal reference scaling using *in silico* references. This normalization is suitable when one client has multiple TMT plexes and is conducted within each cohort. For each TMT plex, an *in silico* reference sample $\mathrm{Ref}_{plex}$ is created taking the mean value for each protein group across all samples in the TMT plex. Then geometric mean for each protein across all client's *in silico* references is computed:

$$\mathrm{GM} = \exp\left(\frac{1}{d}\sum^{d} \log(\mathrm{Ref}_{plex})\right),$$

where $d$ is the total number of plexes in i-th client.

Using $\mathrm{Ref}_{plex}$, the scaling factor $\mathrm{SF}_{plex}$ is calculated for each protein in each TMT plex as the ratio of the geometric mean to the *in silico* reference for that plex:

$$\mathrm{SF}_{plex} = \frac{\mathrm{GM}}{\mathrm{Ref}_{plex}},$$

and normalized intensities are computed by multiplying with the scaling factor.

Both implemented normalization methods should be done on non-log-transformed data, so after this step $log2(x+1)$log-transformation is applied if required by the coordinator in the analysis settings.

To make possible analysis of all protein groups available we used a design matrix mask $D$ for the next steps. The mask has the number of columns equal to the design matrix and rows for each protein group. The mask creation in more detail is described in Supplementary methods.

In the fifth step, for each protein group in $P$, FedProt fits a linear model in a federated fashion, following the approach described in (Karr et al., 2005). Assuming that protein group intensity $Y$ can be modeled as

$$Y = X\beta + \epsilon,$$

where $X$ is the global design matrix and $\epsilon$ is random noise. The coefficients $\beta$ defining the impact of each variable in the design matrix $X$ on the observed intensities can be estimated as

$$\hat{\beta} = (X^T X)^{-1} X^T Y,$$

and the unscaled standard deviations $\mathrm{stdev}_{\mathrm{unscaled}}$ for the coefficients $\hat{\beta}$ can be estimated as

$$\mathrm{stdev}_{\mathrm{unscaled}} = \sqrt{\mathrm{diag}((X^T X)^{-1})}.$$

To avoid sharing $X^i$ and $Y^i$ containing sensitive patient-level data that would be necessary to obtain $X$ and $Y$, $X^T X$ and $X^T Y$ terms of the equation can be computed through summation of local $(X^i)^T X^i$, and $(X^i)^T Y^i$ computed by clients:

$$X^T X = \sum_{i=1}^{k} (X^i)^T X^i,$$

$$X^T Y = \sum_{i=1}^{k} (X^i)^T Y^i.$$

With the exception of rare cases that are separately checked by clients, $(X^i)^T X^i$ and $(X^i)^T Y^i$ do not reveal any patient-level data and can be shared with the server.

On this step design mask $D$ is used to exclude columns and rows corresponding to missing values from $X^T X$ and columns from $X^T Y$. This is necessary to exclude from the calculations for a particular protein the values belonging to a particular column from the design $X$, since these cohorts do not have values (all are NAs). $D$ usage allows us to simulate behavior of the *lmFit* function from the limma R package when working with missing values in data.

To minimize the risk of reconstruction attack, $(X^i)^T X^i$, $(X^i)^T Y^i$ , as well as any local computation result shared with the aggregating server are protected by additive secret sharing (Cramer et al., 2015). Each client generates $n-1$ randomly sampled masks, $r_1$, …, $r_{n-1}$, as equally distributed random values, and calculates the masked data (M - $r_1$ - … - $r_{n-1}$). This noisy data, alongside the masks (n pieces in total), is communicated with other computational parties while ensuring no party receives more than one piece of data from any specific client. The data pieces are encrypted with each party's public key to ensure the data cannot be intercepted. Each party sums the received pieces and sends the results to the Coordinator. The Coordinator gets the summed parts and obtains the global coefficients $\hat{\beta}$. During the additive secure aggregation process the noises will be canceled without noticeable impact on the outcome in comparison with non-secure aggregation while it does not reveal the clients' intermediate results to the Coordinator or any other parties. By increasing the number of parties, the risk of collusion to reconstruct the clients' intermediate results will be further reduced. To simplify the technical aspects of communicating data, FeatureCloud (Matschinske et al., 2023) passes all data through a Relay server which cannot decrypt the data (Figure 1).

Global estimated coefficients $\hat{\beta}$ are shared with the clients, which use it to calculate the local sum of squared errors $SSE^i = \sum_j^{m_i} (y_j^i - \hat{y^i}_j)^2$, where $\hat{y^i}_j$ is the $j$-th component of $\hat{Y^i} = X^i \hat{\beta}$ . The server aggregates local sum of squared errors $SSE^i$ to the global sum of squared errors $SSE$,

$$SSE = \sum_{i=1}^{k} SSE^i,$$

and computes the residual standard deviations $\sigma$

$$\sigma = \sqrt{\frac{SSE}{m-|V|}},$$

where $m = \sum_i^k m^i$ is the number of samples with detected protein group and $|v|$ is the number of variables in the design matrix except masked with the design matrix mask. In this step, a design mask is applied to ensure that missing values are handled correctly.

The sixth step is performed solely on the server side. Given a set of target classes, the contrast is defined as linear combinations of conditions or target classes in a design matrix $X$ and represented as a contrast matrix $K$. The fitting of these contrasts implies the application of the contrast matrix to the regression coefficients $\hat{\beta}$ and covariance matrix of these coefficients $C$ as following:

$$\hat{\beta}' = \hat{\beta}K,$$

$$C' = K^T C K$$

The standard deviations $\mathrm{st.dev}$ for the coefficients is also updated during this step. Specifically, the covariance matrix $C'$ is scaled by its diagonal to become the correlation matrix, on which the Cholesky decomposition is performed, and the result is then used to transform and scale the $\mathrm{st.dev}$ and get the $\mathrm{st.dev}'$. This replicates the implementation of the *contrasts.fit* function from the limma R package (Ritchie et al., 2015)

Further computations performed on the server side replicate *eBayes* from *limma* (Phipson et al., 2016) and require only global $\hat{\beta}'$, $\sigma^2$ and $\mathrm{st.dev}'$. This step starts with moderated t-statistic calculation. For that variance shrinkage is performed to stabilize the variance estimates across genes. Given a vector of sample variances $\sigma^2$ and their associated degrees of freedom $\mathrm{df}_{\mathrm{residual}}$, the empirical Bayes approach fits an F-distribution to estimate the parameters of the prior distribution.

Using the estimated priors (variance $\sigma^2_{\mathrm{prior}}$ and the degrees of freedom $\mathrm{df}_{\mathrm{prior}}$), the posterior variances $\sigma^2_{\mathrm{post}}$ were calculated as a weighted average of the prior and sample variances:

$$\sigma^2_{\mathrm{post}} = \frac{\mathrm{df}_{\mathrm{residual}} \cdot \sigma^2 + \mathrm{df}_{\mathrm{prior}} \cdot \sigma^2_{\mathrm{prior}}}{\mathrm{df}_{\mathrm{residual}} + \mathrm{df}_{\mathrm{prior}}}.$$

After this the moderated t-statistic $t$ and B-statistic $B$ are computed as:

$$t = \frac{\beta'}{\mathrm{st.dev}'} \cdot \frac{1}{\sqrt{\sigma^2_{\mathrm{post}}}},$$

where $\beta$ represents the estimated coefficients (log-fold-changes) and $\mathrm{st.dev}'$ denotes the unscaled standard deviations of the coefficients.

$$B = \log\left(\frac{p}{1-p}\right) - \frac{\log(r)}{2} + k,$$

where $p$ is the proportion of differentially expressed genes, $r$ is the ratio of the variance of the gene to the prior variance, and $k$ is a function involving the moderated t-statistics and the degrees of freedom. The B-statistic represents the log-odds of differential expression.

Finally, the Benjamini-Hochberg procedure (Benjamini and Hochberg, 1995) is applied to compute adjusted p-values. In the result of this step, a feature table which provides moderated t-statistics, log-fold-changes, confidence intervals, and adjusted p-values is generated. The FedProt workflow can be completed after this step without precursor peptide count adjustment.

The last, seventh step replicates *spectraCounteBayes* from *DEqMS* method (Zhu et al., 2020), which estimates different prior variances for proteins quantified by different numbers of peptide precursors and calculates peptide count-adjusted moderated t-statistics and p-values. The server uses the minimal number of quantified precursor peptides across all samples for each protein for estimating the variance of log-counts and fitting a regression model. As a result, the system outputs a final table with statistical measurements for each feature corrected to the number of precursors.

The resulting table is saved on a server and sent to the clients. This approach in FedProt allows us to obtain the same result as what centralized pooled data analysis would yield while implementing strong privacy-preserving measures.

The FedProt user-friendly implementation is accessible as a FeatureCloud App, making privacy-preserving differential protein abundance analysis available to a broad community of biomedical experts.

## Meta-analysis approaches

In order to evaluate FedProt's accuracy in comparison to meta-analyses, we used three classes of meta-analyses: effect size combination methods, p-value combination methods, and non-parametric rank combination methods (Toro-Domínguez et al., 2021; Tseng et al., 2012).

As methods based on integration of p-values we used Fisher's method (Fisher, 1925) and Stouffer's method (Stouffer et al., 1949). In these methods, p-values obtained from each individual analysis can be integrated into a single combined p-value per protein or gene assuming the sum, minimum or maximum of log-transformed p-values from independent studies follow a certain distribution (Toro-Domínguez et al., 2021).

Fisher's method is a classical method (Fisher, 1925; Kaever et al., 2014) in which the meta-p-values are calculated based on a chi-squared distribution. It is a common method for omics data analysis, but it is sensitive to very small p-values (Toro-Domínguez et al., 2021) and treats large and small p-values asymmetrically (Whitlock, 2005). We used the Fisher's method implementation available in the metaVolcanoR package (Prada et al., 2023).

On the other hand, Stouffer's method (Kaever et al., 2014; Stouffer et al., 1949), also known as normal, Z-method, or Z-transform test, integrates p-values but allows for different study weights and has more power and more precision than Fisher's method (Whitlock, 2005). We used the Stouffer's method implementation from the MetaDE package (Wang et al., 2012).

For effect size combination methods, we used the Random Effects Model (REM) (Toro-Domínguez et al., 2021; Tseng et al., 2012) implementation from metaVolcanoR package (Prada et al., 2023). REM takes into account the heterogeneity between studies by adding a between-study variance (Choi et al., 2003). But estimating this variance can be challenging, especially with a small number of studies (cohorts). This is because REM computes p-values using global effect sizes, assuming a normal distribution.

Furthermore, we used the Rank Product method as a representative of non-parametric rank combination methods. RankProd from the RankProd package is a non-parametric rank-based approach, the significance is assessed by a non-parametric permutation test (Hong et al., 2006). While it is non-parametric and doesn't require homogeneity of variances, it might be less powerful than parametric methods when their assumptions are met.

For all chosen meta-analysis methods except REM, global fold-change was calculated as the mean of local fold-changes, producing the same values. Consequently, only Fisher's method and REM results were utilized for the evaluation of log-fold changes.

## Human serum data

### Sample preparation

For FedProt evaluation we were using a TMT dataset of sixty independent human blood serum samples, comprising 30 from patients with primary FSGS and 30 from healthy controls. Written consent for anonymized data retrieval and storage was obtained. The local ethics committee of the Friedrich-Alexander Universität Erlangen-Nürnberg provided approval for the nephrological biobank of the Klinikum Bayreuth (ethic approval code 264_20 B) and the proteomics analysis (ethic approval code 221_20 B). Approval to perform mass spectrometry of serum samples was given by the ethics committee of the Friedrich-Alexander Universität Erlangen-Nürnberg (ethic approval code 182_19 B and 19_182_2-B).

The samples were separated into three groups, each containing 10 healthy and 10 FSGS samples, and were processed independently by different technicians at different days but using the same protocol. Subsequently, the samples were provided to three different LC-MS/MS locations.

### Sample preparation for mass spectrometry

Samples were prepared by three independent scientists applying harmonized protocol. Briefly, 10 µL of the serum samples were loaded onto depletion columns (High-Select Top14 Abundant Protein Depletion Resin, Thermo Fisher Scientific) to deplete the 14 most abundant serum proteins. Thirty micrograms of the filtrates were reduced, alkylated, and digested using LysC followed by trypsin, applying the filter-aided sample preparation (FASP) protocol by Wisniewski (Wiśniewski, 2018). The samples were subsequently desalted using Oasis HLB 96-well µElution Plates (Waters) and reconstituted in 30 µL of 0.1% formic acid containing 3% acetonitrile. The resulting peptide concentrations were determined using a NanoDrop Microvolume Spectrophotometer (Thermo Fisher Scientific). Subsequently, five micrograms of each sample, along with a pooled common reference sample, were labeled using the TMT-11plex kit (Thermo Fisher Scientific). TMT-labeled samples were combined into six sets, each containing five healthy, five FSGS, and one common reference sample and dried in vacuum. Ultimately, the six sample sets were fractionated (High pH reversed phase peptide fractionation kit, Thermo Fisher Scientific) and the fractions were dried in vacuum.

## LC-MS/MS measurement

Mass spectrometry data were acquired in three independent research centers using their preferred instruments and corresponding parameter setup (Table 7).

**Table 7**. LC-MS/MS measurement overview, human serum dataset.

| Center | Sample type | Set-up |
|---|---|---|
| 1 | Human serum | Easy-nLC 1200, QExactive HF |
| 2 | Human serum | Ultimate 3000 nano, Exploris 480 |
| 3 | Human serum | Easy-nLC 1200, Orbitrap Fusion Lumos |

### Mass spectrometry location 1 — QExactive HF

Prior measurement, all samples were dissolved in 0.1 % formic acid injected into an Easy-nLC 1200 coupled to a Q Exactive HF mass spectrometer (both Thermo Fisher Scientific). Samples were loaded onto a 20-cm analytical HPLC column (75 µm ID Pico Tip fused silica emitter, New Objective) packed in-house using ReproSil-Pur C18-AQ 1.9-µm silica beads (Dr. Maisch GmbH) and separated in a 120-min multistep gradient ranging from 10% solvent B to 90% solvent B (0.1 % formic acid in acetonitrile) at a constant flow rate of 200 nL/min. The nano-HPLC column was drawn to a tip of ~10 µm and acted as the electrospray needle of the MS source. Samples were measured in data-dependent mode applying a MS/MS scan to the Top 10 most abundant precursors per survey scan and a dynamic exclusion of 30 s. HCD collision energy was set to 34 % with an isolation width of 0.7 Da. Survey scans were acquired in a scan range of 300–1650 m/z, a mass resolution of 60,000, an AGC target value of $3 \times 10^6$ and a maximum injection time of 50 ms. For MS/MS scans, AGC target and maximum injection time were set to $1 \times 10^5$ and 110 ms, respectively.

### Mass spectrometry location 2 — Exploris480

Samples were dissolved in 0.1% formic acid and analyzed by online C18 nanoHPLC-MS/MS with a system consisting of an Ultimate3000 nano gradient HPLC system (Thermo, Bremen, Germany), and an Exploris480 mass spectrometer (Thermo Fisher Scientific). Fractions were injected onto a cartridge precolumn (300 µm × 5 mm, C18 PepMap, 5 um, 100 A, and eluted via a homemade analytical nano-HPLC column (30 cm × 75 µm; Reprosil-Pur C18-AQ 1.9 µm, 120 A (Dr. Maisch, Ammerbuch, Germany). The gradient was run from 2% to 36% solvent B (20/80/0.1 water/acetonitrile/formic acid (FA) v/v) in 120 min at 250 nl/min. The nano-HPLC column was drawn to a tip of ~10 µm and acted as the electrospray needle of the MS source. The mass spectrometer was operated in data-dependent MS/MS mode with a cycle time of 3 s, with a HCD collision energy at 36% and recording of the MS2 spectrum in the orbitrap, with a quadrupole isolation width of 1.2 Da. In the master scan (MS1) the resolution was 120,000, the scan range 350-1600, at an AGC target of standard maximum fill time of 50 ms. A lock mass correction on the background ion m/z=445.12003 was used. Precursors were dynamically excluded after n=1 with an exclusion duration of 45 s, and with a precursor range of 30 ppm. Charge states 2-5 were included. For MS2 the first mass was set to 110 Da, and the MS2 scan resolution was 45,000 at an AGC target of 200% fill time of 'auto'.

**Mass spectrometry location 3 — Fusion LUMOS**

TMT-labeled peptides were dissolved in 0.1% formic acid and subsequently analyzed by on-line C18 nanoHPLC-MS/MS with a system consisting of an Easy nLC 1200 gradient HPLC system (Thermo, Bremen, Germany), and an Orbitrap Fusion LUMOS mass spectrometer (Thermo). Fractions were injected onto a homemade precolumn (100 µm × 15 mm; Reprosil-Pur C18-AQ 3 µm, Dr. Maisch, Ammerbuch, Germany) and eluted via a homemade analytical nano-HPLC column (30 cm × 75 µm; Reprosil-Pur C18-AQ 1.9 µm). The analytical column temperature was maintained at 50 C with a Sonation PRSO-V2 column oven. The gradient was run from 2% to 36% solvent B (20%/80%/0.1% water/acetonitrile/formic acid (FA) v/v) in 120 min. The nano-HPLC column was drawn to a tip of ∼10 µm and acted as the electrospray needle of the MS source. The mass spectrometer was operated in data-dependent MS/MS mode with a cycle time of 3 s, with a HCD collision energy at 36% and recording of the MS2 spectrum in the orbitrap, with a quadrupole isolation width of 1.2 Da. In the master scan (MS1) the resolution was 120,000, the scan range 350-1600, at an AGC target of 'standard' maximum fill time of 50 ms. A lock mass correction on the background ion m/z=445.12003 was used. Precursors were dynamically excluded after n=1 with an exclusion duration of 45 s, and with a precursor range of 20 ppm. Charge states 2-5 were included. For MS2 the first mass was set to 110 Da, and the MS2 scan resolution was 50,000 at an AGC target of 200% fill time of 50 ms.

## Raw Data Analysis

Raw mass spectra were uniformly preprocessed and quantified using MaxQuant (v 2.4.2) software (Tyanova et al., 2016) separately for each center. The analysis was conducted with default settings unless otherwise specified. Experimentally acquired mass spectra were searched against a human reference proteome (Uniprot, version 2023_05, reviewed/Swiss-Prot entries only, 20,418 protein sequences). Trypsin/P was set as protease (specific mode) allowing for a maximum of two missed cleavages. Carbamidomethylation of cysteine was set as fixed modification whereas oxidation of methionine and acetylation of protein N-terminus was allowed as variable modification. A minimum of two peptides including one unique peptide were required for protein inference controlling the FDR to < 0.05 in target/decoy mode. Match between runs was enabled.

### Data Filtering and Preprocessing

For protein intensity matrices MaxQuant report were independently preprocessed filtering out decoy, contaminant, and modification site-only entries. The column “Majority.protein.IDs” was used for protein group names, and within a group, proteins were sorted alphabetically. This additional sorting allows for a better intersection of independently processed data. Since FedProt uses *in silico* references in the TMT analysis workflow, 6 reference samples were removed from the dataset before analysis.

Protein groups supported by a single peptide were removed during the central DEqMS analysis. Raw intensities were normalized to the median across all data (from *PRONE* R package https://github.com/lisiarend/PRONE), followed by IRS within each center with an *in silico* reference — the mean of all samples for each pool in the center (modified IRS from *PRONE*). Similar filters and normalizations are also implemented in FedProt. Quality control was performed in R, see Supplementary figure S6-B).

## Bacterial dataset creation

### Sample preparation

We evaluated FedProt using a LFQ dataset of 118 samples generated from *Escherichia coli* MG1655 (DSM 18039) cultures. Single colonies from passage three were inoculated in 5 ml either M9 Pyruvate (1x M9 salts, 2 mM $MgSO_4$, 0.1 mM $CaCl_2$, 40 mM sodium pyruvate) or M9 Glucose medium (1x M9 salts, 2 mM $MgSO_4$, 0.1 mM $CaCl_2$, 20 mM glucose) and grown at 37 °C overnight. Of these overnight cultures, 100 µl culture was inoculated in 10 mL fresh M9 pyruvate or M9 glucose medium. Cells were incubated at 37 °C and at 200 rpm and harvested after six hours (M9 glucose) or 12 hours (M9 pyruvate). For cell pellets cells were centrifuged and medium was removed (no further washing step executed). For cell lysates after removal of the medium cells were lysed in 50 µl 100% TFA for 5 minutes at 55 °C, and the solution neutralized with 450 µl 2 M Tris. Samples were shipped on dry ice either as lysates or as cell pellets (Table 8).

Lab A and lab B received cell pellets, while others (C, D, E) received already lysed cells. Each lab received 20 (19 for lab B and C) unique and 4 shared (quality control) samples, 12 (11) samples per condition. One sample was lost during shipment and one more was excluded during quality control of MS results. Each of the four quality control samples were generated by aliquoting one starting sample, i.e. they are technical replicates.

Mass spectrometry data were acquired in five independent research centers using their preferred sample preparation (in case of cell pellets) and MS measurement protocols and instruments (Table 8).

**Table 8**. LC-MS/MS measurement overview, bacterial dataset.

| Lab | Sample type | Set-up |
|---|---|---|
| A | Cell pellet | Evosep One – Exploris 480 |
| B | Cell pellet | nanoElute – timsTOF Pro |
| C | Cell lysate | Ultimate3000 – Orbitrap Fusion Lumos |
| D | Cell lysate | EASY-nLC 1200 – Exploris 480 |
| E | Cell lysate | Ultimate3000 – QE-HFX |

## Sample preparation for mass spectrometry

**Bacterial cell lysis.**

**Lab A.** 50µL LYSE buffer (from iST kit, PreOmics, Martinsried) were added to the bacterial cell pellets and samples were incubated for 10 minutes at 95°C with shaking at 1000 rpm. For lysis, samples were sonicated in a Bioruptor Pico for 10 cycles of 30 seconds on/30 seconds off at 4°C.

**Lab B**. Bacterial cells were lysed according to the SPEED protocol (Doellinger et al., 2020) with further adaptation (Abele et al., 2023). As mentioned above, 50 µl 100% TFA were added to every sample with subsequent incubation at 55°C for 5 minutes. Subsequently, 450 µl 2M Tris was added to the cell lysates to neutralize the sample.

**Labs C-E**. The laboratories C, D, and E used cell lysates prepared following the same protocol as lab B, except that the lysis was performed by lab C and the cell lysates were sent to labs D and E and diluted.

**Protein digestion and peptide purification.**

Each lab used slightly different protocols for protein digestion, peptide purification, and preparation for MS.

**Lab A.** Sample preparation with the iST kit. Protein concentrations of lysates were determined using bicinchoninic acid (BCA) assay (Pierce, #23252). For sample preparation for MS, 50 µg protein per sample was filled up to 50µL with LYSE buffer. Samples were then incubated at 95°C for 10 minutes (1000 rpm). Then the iST protocol was used for all samples according to manufacturer's guidelines (PreOmics GmbH, Martinsried). Briefly, after reduction and alkylation (LYSE buffer), trypsin and Lys-C were added for digestion and samples incubated for 3 hours at 37°C, 500 rpm. Digestion was stopped by adding the STOP buffer. Resulting peptides were cleaned up on the CARTRIDGE and then eluted. The peptide solution was dried in a Concentrator plus (Eppendorf) and the resulting peptide pellet resuspended in 100µL LC-LOAD buffer.

Solid phase extraction using Evotips. For prefractionation on the Evosep One system, 1 µL of resuspended peptides (approx. 0.5µg per Evotip) were loaded on Evotips. Briefly, Evotips were first rinsed with 20µL of solvent B (0.1% FA/ACN, centrifugation at 800g, 60 seconds) and the C18 material conditioned in isopropanol for 30 seconds. The C18 material was equilibrated with 20µL solvent A (0.1% FA/H2O) before sample (diluted in solvent A) was loaded and tips washed two times with solvent A.

**Lab B.** In-solution tryptic digestion. Protein concentrations of cell lysates were determined using bicinchoninic acid (BCA) assay (Interchim Uptima, Paris, France). 10 µg of protein were used for tryptic digestion. Proteins were reduced and alkylated by additions of 100mM Tris(2-carboxyethyl)phosphine (TCEP) and 440 mM chloroacetic acid (CAA). Samples, covered with aluminum foil, were then incubated at 95°C (5 minutes, 400 rpm). Next, samples were diluted from 2 M to 1M Tris and cooled. Finally, overnight proteolytic digestion at 37°C (400 rpm) was performed by adding trypsin at a protease to protein ratio of 1:50.

Solid phase extraction using StageTips. Samples were acidified to pH < 3. In-house built C18 Stage Tips were equilibrated with 250 µl 100% ACN and washed with 250 ul elution solution (40% ACN, 0.1% FA) followed by 250 µl washing solution (0.1% FA). Next, the digests were loaded onto the column and the stage tips were centrifuged and washed with 250 µl washing solution (0.1% FA). Finally, peptides were eluted twice with 40 µl elution solution (40% ACN, 0.1% FA). Samples were dried in a SpeedVac at 35°C and reconstituted in 50 µl 0.1% FA.

**Lab C.** In-solution Tryptic Digestion. This step was performed similarly to Lab B, but with variations in TCEP (9 mM) and CAA (33 mM) concentrations, and a heated cap was used during incubation.

Solid Phase Extraction using StageTips (Rappsilber et al., 2007). Samples were acidified (pH < 3 with 6% formic acid). The in-house build C18 StageTips, with three Empore C18 (3M) disks, were equilibrated consecutively with 250 µl 100% ACN, 250 ul elution solution (40% ACN, 0.1% FA) and 250 µl washing solution (2% ACN, 0.1% FA; each step: 2 min at 211 rcf). Next, the digested sample was loaded onto the StageTip, centrifuged (5 min at 211 rcf), and afterwards washed (2% ACN, 0.1% FA; 2 min at 211 rcf). Finally, peptides were eluted twice with 50 µl elution solution (40% ACN, 0.1% FA; 2 minutes at 500g). All samples were dried in a centrifugal evaporator (Centrivap Cold Trap -50, Labconco, US) and stored at -80°C.

**Lab D.** Protein Digestion. Proteins were reduced and alkylated using 4 µL of 100 mM TCEP and CAA, respectively, and incubated at 95°C, 400 rpm, 5 minutes. Lysate dilution was to 1M Tris using ddH$_2$O. Trypsin digestion (0.2 µg, 1:50 ratio) was done by overnight incubation at 37°C at 400 rpm. The digest was quenched at a final concentration of 3 % FA.

Solid phase extraction using StageTips. Peptides were purified using the in-StageTip protocol (Rappsilber et al., 2007) and styrenedivinylbenzene reverse-phase sulfonate (SDB-RPS, Empore™ SPE Disks, CDS Analytical, 98-0604-0226-4). In brief, a total of 20 µg of peptides was loaded on the stage tips (500xg, 10 minutes). Peptides were washed twice using 1% TFA (v/v) in isopropanol and once using 0.2% TFA in MS-grade H2O (1000xg). Peptides were then eluted (80% acetonitrile (v/v), 1% NH4+ (v/v) in MS-grade H$_2$O) at 300xg and dried (60 minutes, 45°C, SpeedVac centrifuge, Eppendorf). Samples were resuspended (0.1% (v/v) TFA, 2% acetonitrile in MS-grade H$_2$O) and stored at -20°C.

**Lab E.** In-solution Tryptic Digestion was performed using the same protocol as in Lab B.

Solid Phase Extraction Using SDBRPS StageTips (Rappsilber et al., 2007). Samples were acidified to a pH<3 with FA. The in-house build SDBRPS StageTips using two Empore SDBRPS (3M) disks were equilibrated consecutively with 100 µl 100% ACN, 100 µl 30% MeOH, 1% TFA and 150 µl washing solution 3 (0.2% TFA; each step: 1 min at 800g). Next, the digested sample was diluted in 200µl 1% TFA and loaded onto the StageTip (1 min, 800g). Then StageTips were washed three times. With 100µl washing solution 1 (99% ethyl acetate and 1% TFA, 1 min, 800g); with 100µl washing solution 2 (99% isopropanol and 1% TFA, 1 min, 800g), with 150µl washing solution 3 (2 min at 800g). Finally, peptides were eluted with 60 µl elution solution (80% ACN, 5% from 25% $NH_4OH$. 2 min, 800g). Post drying, samples were stored at -20°C.

## LC-MS/MS measurement

**Lab A.** Samples were analyzed in a randomized injection manner on an Evosep One LC using the 30 spd (samples per day) method using a 15cm x 150µm x 1.5µm column from PepSep heated to 40°C in a column oven (Sonation GmbH). Eluted peptides were electrosprayed into an Exploris 480 mass spectrometer (Thermo Fisher Scientific, Bremen). The MS was operated in an data-independent acquisition mode. MS1 spectra (380–980 m/z) were recorded at a resolution of 120,000 using an automatic gain control (AGC) target value of 300% and maximum injection time of 100 ms. MS2 spectra were acquired at a resolution of 30,000, with an automatic gain control (AGC) target value of 3000% and auto maximum injection time. Isolation windows were 20 m/z with an overlap of 1 m/z, resulting in 30 windows. Normalized collision energy was set to 30% and data was acquired in centroid mode.

**Lab B**. Samples (3 µl, around 300 ng of peptides) were analyzed on a nanoElute LC coupled to a timsTOF Pro mass spectrometer with a CaptiveSpray ion source (Bruker, Germany). Samples were injected on a self-packed C18 column (75µm internal diameter) with 1.9 µm ReproSil-Pur 120 C18-AQ resin (Dr Maisch, Germany). A gradient of water (A) and acetonitrile (B) supplemented with 0.1% formic acid was applied at a flow rate of 300 nL/min and a column temperature of 50°C. The following gradient was applied: 0 min, 2%B; 2 min, 5%B; 62 min, 24%B; 72 min, 35%B; 75 min, 60%B; 78 min, 85%B. The MS was operated in a data-independent acquisition parallel accumulation-serial fragmentation (PASEF) mode. Ion accumulation and separation using trapped ion mobility spectrometry (TIMS) was set to a ramp time of 100 ms. One scan cycle included one TIMS full MS scan and two rows of 30 windows with a width of 25 m/z covering a range of 400-1,150 m/z. 5 scans per PASEF scan were applied.

**Lab C.** Around 500 ng peptides dissolved in washing solution (2% ACN, 0.1% FA) were analyzed on a Dionex Ultimate 3000 RSLCnano system coupled to an Orbitrap Fusion Lumos Tribrid Mass Spectrometer (Thermofisher Scientific, Bremen). Injected peptides were delivered to a trap column (ReproSil-pur C18-AQ, 5 µm, Dr. Maisch, 20 mm × 75 µm, self-packed) at a flow rate of 5 µL/min in 100% solvent A (0.1% formic acid in HPLC grade water). After 10 min of loading, peptides were transferred to an analytical column (ReproSil Gold C18-AQ, 3 µm, Dr. Maisch GmbH, 400 mm × 75 µm, self-packed) and separated using a 60 min linear gradient from 4% to 34% of solvent B (0.1% formic acid in acetonitrile and 5% (v/v) DMSO) at 300 nL/min flow rate. Both nanoLC solvents contained 5% (v/v) DMSO. The Fusion Lumos Tribrid Mass Spectrometer mass spectrometer was operated in data independent acquisition and positive ionization mode. MS1 spectra (360–1300 m/z) were recorded at a resolution of 60,000 using an automatic gain control (AGC) target value of 1e6 and maximum injection time (maxIT) of 50 ms. MS2 spectra were acquired at a resolution of 30,000, a scan range of 200-1,800 m/z, and with an automatic gain control (AGC) target value of 5E5 and maximum injection time (maxIT) of 54 ms. We used a variable window acquisition scheme with 40 windows overlapping by 1 m/z (see Supplementary Table S3) with a default charge state of two. Fragmentation was performed using higher energy collision induced dissociation (HCD) and a normalized collision energy of 30%.

**Lab D.** MS data were acquired on an EASY-nLC 1200 ultrahigh-pressure system (Thermo Fisher Scientific, San Jose, USA) coupled to an Orbitrap Exploris 480 Mass Spectrometer (Thermo Fisher Scientific, Waltham, USA) using a nano-electrospray ion source (Thermo Fisher Scientific). A total of 200 ng peptides was injected into a 50 cm-column (inner diameter: 75µm, generated in-house (Müller-Reif et al., 2021) using ReproSil-Pur C18-AQ 1.9µm beads from Dr. Maisch GmbH, Ammerbuch, Germany). The temperature was kept constant at 55°C in a column oven. A two-buffer system enabled the gradual elution of peptides: buffer A (0.1% FA in $H_2O$) and buffer B (80% acetonitrile, 0.1 % FA in $H_2O$). During the course of liquid chromatography (LC), buffer B was increased from 2% to 35% within the first 60 minutes, followed by a further increase to 50% within 6 min, to 60% within 4 min and to 90% within 1 min which was kept constant for 4 min to ensure a complete elution of peptides. The flow rate was kept constant at 300 nl/min. DIA of the MS experiments included MS1 scans (scan range: 300 to 1,650 m/z; resolution: 120,000; maximum injection time: 20 ms; normalized AGC target: 300%) as well as sequential MS2 scans (resolution: 30,000; maximum injection time: 54 ms; normalized AGC target: 3000%) using 44 DIA isolation windows. Peptides were fragmented using stepped HCD collision energies (25, 27.5, 30).

**Lab E.** The MS data were acquired in DIA mode on a QExactive-HFX mass spectrometer (Thermo Fisher Scientific Inc., Waltham, MA, USA). Around 400 ng per sample were automatically loaded to the online coupled RSLC (Ultimate 3000, Thermo Fisher Scientific Inc.) HPLC system. A Nano-Trap column was used (300-µm inner diameter (ID) × 5 mm, packed with Acclaim PepMap100 C18, 5µm, 100 Å from LC Packings, Sunnyvale, CA, USA), before separation by reversed-phase chromatography (Acquity UPLC M-Class HSS T3 Column 75µm ID × 250 mm, 1.8µm from Waters, Eschborn, Germany) at 40°C. Peptides were eluted from the column at 250 nl/min using increasing ACN concentration in 0.1% formic acid from 3 to 40% over a 95-min gradient. The DIA method consisted of a survey scan from 300 to 1,500 m/z at 120,000 resolution and an automatic gain control (AGC) target of 3E6 or 120-ms maximum injection time. Fragmentation was performed via higher-energy collisional dissociation with a target value of 3E6 ions determined with predictive AGC. Precursor peptides were isolated with 37 variable windows spanning from 300 to 1,650 m/z at 30,000 resolution with an AGC target of 3E6 and automatic injection time. The normalized collision energy was 28, and the spectra were recorded in profile type.

## Raw Data Analysis

Raw mass spectra were uniformly preprocessed and quantified using DIA-NN (Demichev et al., 2020) (https://github.com/vdemichev/DIA-NN), v.1.8.1 in a separate run for each lab. The analysis was conducted with default settings unless otherwise specified. An *in silico* spectral library was generated in each DIA-NN run from *E. coli* MG1655 (taxID 83333) protein sequence database (Uniprot UP000000625, 4448 entries).

The analysis was carried out in “any LC (high-accuracy)” mode. Two missed cleavage and a maximum of two variable modifications per peptide were allowed: acetylation of protein N-termini and oxidation of methionine. Min precursor m/z was set to 360. The match-between-runs was enabled. The data was reanalyzed utilizing a deep-learning generated spectral library to refine the results. For specific parameters on setup of the DIA-NN searches, see Supplementary Table S4.

## Data Filtering and Preprocessing

Protein quantities were obtained from the MaxLFQ (Cox et al., 2014) algorithm implemented in DIA-NN v 1.8.1. For protein intensity matrices, DIA-NN outputs were filtered using the criteria: Lib.Q.Value ≤ 0.01 and Lib.PG.Q.Value ≤ 0.01. Since MaxLFQ protein quantities are already normalized, no additional normalization was executed either during preprocessing or implemented in FedProt.

Quality control was performed in R (Supplementary Figure S1, S6), one sample from Lab C's dataset was excluded after quality control due to being an outlier (Supplementary Figure S9).

# Simulated data

To generate simulated data we used the approach proposed by Wang et al. (Wang et al., 2021):

$$y_{pj} \sim \begin{cases} \mathcal{N}(\mu_p, \sigma_p^2) & \text{w.p. } (1-\pi_p) \\ \mathcal{N}(\mu_p + \Delta\mu, \sigma_p^2) & \text{w.p. } \pi_p \end{cases}$$

where $y_{pj}$ represents the intensity for $p$-th protein, $p = 1, \ldots, n$, from $j$-th sample, $j = 1, \ldots, m$.

Protein intensity $y_{pj}$ is modeled from mixture distribution, where $\pi_p \in [0, 0.5)$ is the outlier proportion. Outliers could be differentially abundant proteins or technical errors. We didn’t add a sample effect to our model because we simulated data after the MaxLFQ method, which contains the normalization step eliminating it (Cox et al., 2014). The protein population distribution parameters were generated separately for each protein, means $\mu_p$ were from $\mathcal{N}(0, 2)$ and variances $\sigma_p^2$ were from an Inverse Gamma distribution.

We adapted the *sim.dat.fn* function from the RobNorm package (Wang et al., 2021), with modifications to the inverse gamma distribution parameters (the shape parameter of 2 and scale parameter of 3). Parameter $\Delta\mu$ was used to represent a shift in a differentially regulated block, to generate up- or down-regulated proteins. The protein in the block has a chance, derived from a Binomial distribution with a success probability of 0.8, of undergoing a shift. $\Delta\mu$ was used to represent a shift in a differentially regulated block, to generate up- or down-regulated proteins.

We generated the data without batch effects first, with 6000 proteins and 600 samples, 300 each in condition A and B. The block consisting of 200 proteins differentially abundant between conditions A and B was obtained using $\Delta\mu = 1.25$. Additionally, to simulate the effects of unknown covariate, a block of 150 proteins each were generated, with $\Delta\mu = 1.25$, and randomly assigned to samples in class B. The proportion of samples in class B with unknown covariate is shown in Table 4.

To simulate a multi-center study, we then randomly split the data into three cohorts and added batch effects. To investigate the effect of data imbalance on method performances, data were splitted into cohorts twice, once such that cohort sizes and condition A and B frequencies were equal and once unequal across cohorts (see Table 4 for details).

To simulate batch effects in our data we utilized the ComBat model (Johnson et al., 2007) designed for removing batch effects:

$$y_{pji} = y_{pj} + \gamma_{pi} + \delta_{pi}\epsilon_{pji}$$

where $i = 1, \ldots, k$, and $k$ is the total number of batches, errors $\epsilon_{pji}$ are normally distributed $N(0, 1)$, additive batch parameter $\gamma_{pi}$ drawn from Normal distribution and multiplicative batch parameter $\delta_{pi}$ drawn from Inverse Gamma distribution. For simulation we used $\gamma_{p1} \sim N(0, 1)$ for additive batch effect and $\delta_{p1} \sim IG(3, 2)$ for multiplicative batch effect for the first batch, $\gamma_{p2} \sim N(0.2, 0.5)$ and $\delta_{p2} \sim IG(2.5, 1)$ for the second batch, and $\gamma_{p3} \sim N(-0.2, 1.5)$ and $\delta_{p3} \sim IG(4, 0.5)$ for the third batch.

Missing values were introduced to the data using the approach described previously (Jin et al., 2021), with 0.2 for missing values rate and 0.5 for missing not at random rate, to the total of 20% of missing values in the dataset.

## Data analysis

Data after quantification and preprocessing were analyzed in R. Filtering based on the number of missing values per class was done using the 80% threshold, the same as FedProt default value. For the TMT dataset median normalization from the *PRONE* package was used (https://github.com/lisiarend/PRONE). For the meta-analyses we used *MetaDE*, *MetaVolcanoR*, *RankProd* R packages.

Adjusted p-values used for evaluation using the bacterial and human serum datasets are count-scaled BH-method adjusted p-values, similar to what DEqMS is calculating. Before log-transformation of adjusted p-values, a small value (1E-300) was used to replace 0 in REM results for the bacterial dataset. For evaluation using the simulated dataset, BH-adjusted p-values were not scaled using counts, because we did not simulate spectra counts data.

Evaluation was done using Python. For upset plots python *upsetplot* library was used (https://upsetplot.readthedocs.io/en/stable/). For other plots — matplotlib and seaborn packages.

To investigate the impact of including batch effects in the design, the data were preprocessed in the same way, depending on the dataset. The difference was the incorporation of the batches into the design in the analysis, or the correction of the data before (after preprocessing) and the analysis without this information in the design.

# Data availability

The mass spectrometry proteomics data have been deposited to the ProteomeXchange Consortium via the PRIDE (Perez-Riverol et al., 2022) partner repository with the dataset identifiers PXD053812 (the bacterial dataset) and PXD053560 (the human serum dataset).

# Code availability

The code used for the data preprocessing, quality control, and the simulated dataset generation, together with the code for running FedProt are available at GitHub (https://github.com/Freddsle/FedProt) and FeatureCloud App Store.

# Acknowledgements

The authors acknowledge Fabian Gruhn and Ulrike Scholz for technical assistance. Figure 1 was created with BioRender.com.

# Funding

This work was supported by the German Federal Ministry of Education and Research (BMBF) within the framework of "CLINSPECT-M/-2" (grant 161L0214A, 16LW0243K and 16LW0248). This work was supported by the German Federal Ministry of Education and Research (BMBF) within the framework of the *e:Med* research and funding concept (*grants 01ZX1910D and 01ZX2210D*). This work was supported by the German Federal Ministry of Education and Research (BMBF) within the framework of the *e:Med* research and funding concept, project "SyMBoD" (*grants 01ZX2210B and 01ZX1910B*). This project has received funding from the European Union's Horizon2020 research and innovation programme under grant agreement No 826078. This publication reflects only the authors' view and the European Commission is not responsible for any use that may be made of the information it contains. This work was developed as part of the FeMAI project and is funded by the German Federal Ministry of Education and Research (BMBF) under grant number 01IS21079. This work was funded by the German Federal Ministry of Education and Research (BMBF) (*grant 16DTM100A*). This work was funded by the Deutsche Forschungsgemeinschaft (DFG, German Research Foundation) under Germany’s Excellence Strategy within the framework of the Munich Cluster for Systems Neurology (EXC 2145 SyNergy – ID 390857198). T.L. was awarded with a seed funding under the Excellence Strategy of the Federal Government and the Länder.

# Contributions

Y.B., M.B., C.M., V.S., O.Z. — FedProt differential abundance analysis algorithm development. M.B., A.H., T.F., J.M., J.S., R.R. — FedProt privacy-preserving measures development. Y.B., M.B., O.Z. — FedProt implementation and testing. M.A. — Bacterial dataset generation. Y.B., M.A., C. von T., T.B., L.S., P.G., S.M.H., S.L., A.I., M.M., C.L., B.K. — Bacterial dataset measurement, quantification and preprocessing. J.R.S., S.K., J.M.-D., P. A van V., Y.M. — Human serum dataset generation, measurement and quantification. Y.B., L.A., T.L. — Human serum dataset preprocessing. Y.B., L.A., O.Z. — Simulated data generation. Y.B., E.H., L.A., K.A., T.L., O.Z. — Data analysis and interpretation. Y.B., M.A., M.B., C. von T., T.B., L.S., P.G., J.R.S., C.L., J.B., O.Z. — Drafted the manuscript and designed the figures (with input from all authors). J.R.S., S.K., C.L., B.K., J.B., O.Z. — designed the study and supervised the work. All authors have read and approved the final version of the manuscript.

# References

Abele, M., Doll, E., Bayer, F.P., Meng, C., Lomp, N., Neuhaus, K., Scherer, S., Kuster, B., Ludwig, C., 2023. Unified Workflow for the Rapid and In-Depth Characterization of Bacterial Proteomes. Mol. Cell. Proteomics 22. https://doi.org/10.1016/j.mcpro.2023.100612

Adamowicz, K., Arend, L., Maier, A., Schmidt, J.R., Kuster, B., Tsoy, O., Zolotareva, O., Baumbach, J., Laske, T., 2023. Proteomic meta-study harmonization, mechanotyping and drug repurposing candidate prediction with ProHarMeD. Npj Syst. Biol. Appl. 9, 1–10. https://doi.org/10.1038/s41540-023-00311-7

Aebersold, R., Mann, M., 2016. Mass-spectrometric exploration of proteome structure and function. Nature 537, 347–355. https://doi.org/10.1038/nature19949

Aljawad, M.F., Faisal, A.H.M.A., Alqanbar, M.F., Wilmarth, P.A., Hassan, B.Q., 2023. Tandem mass tag-based quantitative proteomic analysis of cervical cancer. PROTEOMICS – Clin. Appl. 17, 2100105. https://doi.org/10.1002/prca.202100105

Altelaar, A.F.M., Munoz, J., Heck, A.J.R., 2013. Next-generation proteomics: towards an integrative view of proteome dynamics. Nat. Rev. Genet. 14, 35–48. https://doi.org/10.1038/nrg3356

Benjamini, Y., Hochberg, Y., 1995. Controlling the False Discovery Rate: A Practical and Powerful Approach to Multiple Testing. J. R. Stat. Soc. Ser. B Methodol. 57, 289–300. https://doi.org/10.1111/j.2517-6161.1995.tb02031.x

Breitling, R., Armengaud, P., Amtmann, A., Herzyk, P., 2004. Rank products: a simple, yet powerful, new method to detect differentially regulated genes in replicated microarray experiments. FEBS Lett. 573, 83–92. https://doi.org/10.1016/j.febslet.2004.07.055

Brenes, A., Hukelmann, J., Bensaddek, D., Lamond, A.I., 2019. Multibatch TMT Reveals False Positives, Batch Effects and Missing Values*. Mol. Cell. Proteomics 18, 1967–1980. https://doi.org/10.1074/mcp.RA119.001472

Bruderer, R., Bernhardt, O.M., Gandhi, T., Xuan, Y., Sondermann, J., Schmidt, M., Gomez-Varela, D., Reiter, L., 2017. Optimization of Experimental Parameters in Data-Independent Mass Spectrometry Significantly Increases Depth and Reproducibility of Results*. Mol. Cell. Proteomics 16, 2296–2309. https://doi.org/10.1074/mcp.RA117.000314

Bullard, J.H., Purdom, E., Hansen, K.D., Dudoit, S., 2010. Evaluation of statistical methods for normalization and differential expression in mRNA-Seq experiments. BMC Bioinformatics 11, 1–13.
Choi, J.K., Yu, U., Kim, S., Yoo, O.J., 2003. Combining multiple microarray studies and modeling interstudy variation. Bioinformatics 19, i84–i90.
Collins, B.C., Hunter, C.L., Liu, Y., Schilling, B., Rosenberger, G., Bader, S.L., Chan, D.W., Gibson, B.W., Gingras, A.-C., Held, J.M., 2017. Multi-laboratory assessment of reproducibility, qualitative and quantitative performance of SWATH-mass spectrometry. Nat. Commun. 8, 291.
Cox, J., Hein, M.Y., Luber, C.A., Paron, I., Nagaraj, N., Mann, M., 2014. Accurate Proteome-wide Label-free Quantification by Delayed Normalization and Maximal Peptide Ratio Extraction, Termed MaxLFQ*. Mol. Cell. Proteomics 13, 2513–2526. https://doi.org/10.1074/mcp.M113.031591
Cramer, R., Damgård, I.B., Nielsen (aut), J.B., 2015. Secure Multiparty Computation. Cambridge University Press.
Čuklina, J., Lee, C.H., Williams, E.G., Sajic, T., Collins, B.C., Rodríguez Martínez, M., Sharma, V.S., Wendt, F., Goetze, S., Keele, G.R., Wollscheid, B., Aebersold, R., Pedrioli, P.G.A., 2021. Diagnostics and correction of batch effects in large-scale proteomic studies: a tutorial. Mol. Syst. Biol. 17, e10240. https://doi.org/10.15252/msb.202110240
Demichev, V., Messner, C.B., Vernardis, S.I., Lilley, K.S., Ralser, M., 2020. DIA-NN: neural networks and interference correction enable deep proteome coverage in high throughput. Nat. Methods 17, 41–44. https://doi.org/10.1038/s41592-019-0638-x
Doellinger, J., Schneider, A., Hoeller, M., Lasch, P., 2020. Sample Preparation by Easy Extraction and Digestion (SPEED) - A Universal, Rapid, and Detergent-free Protocol for Proteomics Based on Acid Extraction *. Mol. Cell. Proteomics 19, 209–222. https://doi.org/10.1074/mcp.TIR119.001616
Fierro-Monti, I., Wright, J.C., Choudhary, J.S., Vizcaíno, J.A., 2022. Identifying individuals using proteomics: are we there yet? Front. Mol. Biosci. 9.
Fisher, R.A., 1925. Statistical Methods for Research Workers (Oliver and Boyd, Edinburgh, UK).
Fröhlich, K., Fahrner, M., Brombacher, E., Seredynska, A., Maldacker, M., Kreutz, C., Schmidt, A., Schilling, O., 2024. Data-independent acquisition: A milestone and prospect in clinical mass spectrometry-based proteomics. Mol. Cell. Proteomics 0. https://doi.org/10.1016/j.mcpro.2024.100800
Geyer, P.E., Mann, S.P., Treit, P.V., Mann, M., 2021. Plasma Proteomes Can Be Reidentifiable and Potentially Contain Personally Sensitive and Incidental Findings. Mol. Cell. Proteomics 20. https://doi.org/10.1074/mcp.RA120.002359
Haidich, A.B., 2010. Meta-analysis in medical research. Hippokratia 14, 29–37.
Hernández, B., Parnell, A., Pennington, S.R., 2014. Why have so few proteomic biomarkers “survived” validation? (Sample size and independent validation considerations). PROTEOMICS 14, 1587–1592. https://doi.org/10.1002/pmic.201300377
Higgins, J.P.T., Thompson, S.G., 2002. Quantifying heterogeneity in a meta-analysis. Stat. Med. 21, 1539–1558. https://doi.org/10.1002/sim.1186
Hong, F., Breitling, R., McEntee, C.W., Wittner, B.S., Nemhauser, J.L., Chory, J., 2006. RankProd: a bioconductor package for detecting differentially expressed genes in meta-analysis. Bioinformatics 22, 2825–2827.
Jin, L., Bi, Y., Hu, C., Qu, J., Shen, S., Wang, X., Tian, Y., 2021. A comparative study of evaluating missing value imputation methods in label-free proteomics. Sci. Rep. 11, 1760. https://doi.org/10.1038/s41598-021-81279-4

Johnson, W.E., Li, C., Rabinovic, A., 2007. Adjusting batch effects in microarray expression data using empirical Bayes methods. Biostatistics 8, 118–127. https://doi.org/10.1093/biostatistics/kxj037

Kaever, A., Landesfeind, M., Feussner, K., Morgenstern, B., Feussner, I., Meinicke, P., 2014. Meta-analysis of pathway enrichment: combining independent and dependent omics data sets. PLoS One 9, e89297.

Karr, A.F., Lin, X., Sanil, A.P., Reiter, J.P., 2005. Secure Regression on Distributed Databases. J. Comput. Graph. Stat. 14, 263–279. https://doi.org/10.1198/106186005X47714

Law, C.W., Alhamdoosh, M., Su, S., Dong, X., Tian, L., Smyth, G.K., Ritchie, M.E., 2016. RNA-seq analysis is easy as 1-2-3 with limma, Glimma and edgeR. F1000Research 5, ISCB Comm J-1408. https://doi.org/10.12688/f1000research.9005.3

Lazar, C., Gatto, L., Ferro, M., Bruley, C., Burger, T., 2016. Accounting for the Multiple Natures of Missing Values in Label-Free Quantitative Proteomics Data Sets to Compare Imputation Strategies. J. Proteome Res. 15, 1116–1125. https://doi.org/10.1021/acs.jproteome.5b00981

Lehmann, B.D., Colaprico, A., Silva, T.C., Chen, J., An, H., Ban, Y., Huang, H., Wang, L., James, J.L., Balko, J.M., Gonzalez-Ericsson, P.I., Sanders, M.E., Zhang, B., Pietenpol, J.A., Chen, X.S., 2021. Multi-omics analysis identifies therapeutic vulnerabilities in triple-negative breast cancer subtypes. Nat. Commun. 12, 6276. https://doi.org/10.1038/s41467-021-26502-6

Ludwig, C., Gillet, L., Rosenberger, G., Amon, S., Collins, B.C., Aebersold, R., 2018. Data-independent acquisition-based SWATH-MS for quantitative proteomics: a tutorial. Mol. Syst. Biol. 14, e8126. https://doi.org/10.15252/msb.20178126

Makinde, F.L., Tchamga, M.S., Jafali, J., Fatumo, S., Chimusa, E.R., Mulder, N., Mazandu, G.K., 2021. Reviewing and assessing existing meta-analysis models and tools. Brief. Bioinform. 22, bbab324.

Matschinske, J., Späth, J., Bakhtiari, M., Probul, N., Kazemi Majdabadi, M.M., Nasirigerdeh, R., Torkzadehmahani, R., Hartebrodt, A., Orban, B.-A., Fejér, S.-J., Zolotareva, O., Das, S., Baumbach, L., Pauling, J.K., Tomašević, O., Bihari, B., Bloice, M., Donner, N.C., Fdhila, W., Frisch, T., Hauschild, A.-C., Heider, D., Holzinger, A., Hötzendorfer, W., Hospes, J., Kacprowski, T., Kastelitz, M., List, M., Mayer, R., Moga, M., Müller, H., Pustozerova, A., Röttger, R., Saak, C.C., Saranti, A., Schmidt, H.H.H.W., Tschohl, C., Wenke, N.K., Baumbach, J., 2023. The FeatureCloud Platform for Federated Learning in Biomedicine: Unified Approach. J. Med. Internet Res. 25, e42621. https://doi.org/10.2196/42621

McMahan, B., Moore, E., Ramage, D., Hampson, S., Arcas, B.A. y, 2017. Communication-Efficient Learning of Deep Networks from Decentralized Data, in: Proceedings of the 20th International Conference on Artificial Intelligence and Statistics. Presented at the Artificial Intelligence and Statistics, PMLR, pp. 1273–1282.

Melis, L., Song, C., De Cristofaro, E., Shmatikov, V., 2019. Exploiting Unintended Feature Leakage in Collaborative Learning, in: 2019 IEEE Symposium on Security and Privacy (SP). Presented at the 2019 IEEE Symposium on Security and Privacy (SP), pp. 691–706. https://doi.org/10.1109/SP.2019.00029

Müller-Reif, J.B., Hansen, F.M., Schweizer, L., Treit, P.V., Geyer, P.E., Mann, M., 2021. A New Parallel High-Pressure Packing System Enables Rapid Multiplexed Production of Capillary Columns. Mol. Cell. Proteomics 20. https://doi.org/10.1016/j.mcpro.2021.100082

Muntel, J., Gandhi, T., Verbeke, L., M. Bernhardt, O., Treiber, T., Bruderer, R., Reiter,

L., 2019. Surpassing 10000 identified and quantified proteins in a single run by optimizing current LC-MS instrumentation and data analysis strategy. Mol. Omics 15, 348–360. https://doi.org/10.1039/C9MO00082H
Myint, L., Kleensang, A., Zhao, L., Hartung, T., Hansen, K.D., 2017. Joint Bounding of Peaks Across Samples Improves Differential Analysis in Mass Spectrometry-Based Metabolomics. Anal. Chem. 89, 3517–3523. https://doi.org/10.1021/acs.analchem.6b04719
Nasirigerdeh, R., Torkzadehmahani, R., Baumbach, J., Blumenthal, D.B., 2021. On the Privacy of Federated Pipelines, in: Proceedings of the 44th International ACM SIGIR Conference on Research and Development in Information Retrieval. Presented at the SIGIR '21: The 44th International ACM SIGIR Conference on Research and Development in Information Retrieval, ACM, Virtual Event Canada, pp. 1975–1979. https://doi.org/10.1145/3404835.3462996
Nygaard, V., Rødland, E.A., Hovig, E., 2016. Methods that remove batch effects while retaining group differences may lead to exaggerated confidence in downstream analyses. Biostat. Oxf. Engl. 17, 29–39. https://doi.org/10.1093/biostatistics/kxv027
Perez-Riverol, Y., Bai, J., Bandla, C., García-Seisdedos, D., Hewapathirana, S., Kamatchinathan, S., Kundu, D.J., Prakash, A., Frericks-Zipper, A., Eisenacher, M., 2022. The PRIDE database resources in 2022: a hub for mass spectrometry-based proteomics evidences. Nucleic Acids Res. 50, D543–D552.
Phipson, B., Lee, S., Majewski, I.J., Alexander, W.S., Smyth, G.K., 2016. ROBUST HYPERPARAMETER ESTIMATION PROTECTS AGAINST HYPERVARIABLE GENES AND IMPROVES POWER TO DETECT DIFFERENTIAL EXPRESSION. Ann. Appl. Stat. 10, 946–963. https://doi.org/10.1214/16-AOAS920
Prada, C., Lima, D., Nakaya, H., 2023. MetaVolcanoR: Gene Expression Meta-analysis Visualization Tool. https://doi.org/10.18129/B9.bioc.MetaVolcanoR
Rappsilber, J., Mann, M., Ishihama, Y., 2007. Protocol for micro-purification, enrichment, pre-fractionation and storage of peptides for proteomics using StageTips. Nat. Protoc. 2, 1896–1906. https://doi.org/10.1038/nprot.2007.261
Ritchie, M.E., Phipson, B., Wu, D., Hu, Y., Law, C.W., Shi, W., Smyth, G.K., 2015. limma powers differential expression analyses for RNA-sequencing and microarray studies. Nucleic Acids Res. 43, e47. https://doi.org/10.1093/nar/gkv007
Rozanova, S., Barkovits, K., Nikolov, M., Schmidt, C., Urlaub, H., Marcus, K., 2021. Quantitative Mass Spectrometry-Based Proteomics: An Overview, in: Marcus, K., Eisenacher, M., Sitek, B. (Eds.), Quantitative Methods in Proteomics, Methods in Molecular Biology. Springer US, New York, NY, pp. 85–116. https://doi.org/10.1007/978-1-0716-1024-4_8
Sajic, T., Liu, Y., Aebersold, R., 2015. Using data-independent, high-resolution mass spectrometry in protein biomarker research: Perspectives and clinical applications. PROTEOMICS – Clin. Appl. 9, 307–321. https://doi.org/10.1002/prca.201400117
Stouffer, S.A., Suchman, E.A., DeVinney, L.C., Star, S.A., Williams Jr, R.M., 1949. The american soldier: Adjustment during army life.(studies in social psychology in world war ii), vol. 1.
Toro-Domínguez, D., Villatoro-García, J.A., Martorell-Marugán, J., Román-Montoya, Y., Alarcón-Riquelme, M.E., Carmona-Sáez, P., 2021. A survey of gene expression meta-analysis: methods and applications. Brief. Bioinform. 22, 1694–1705. https://doi.org/10.1093/bib/bbaa019
Tseng, G.C., Ghosh, D., Feingold, E., 2012. Comprehensive literature review and statistical considerations for microarray meta-analysis. Nucleic Acids Res. 40, 3785–3799. https://doi.org/10.1093/nar/gkr1265

Turner, R.M., Bird, S.M., Higgins, J.P., 2013. The impact of study size on meta-analyses: examination of underpowered studies in Cochrane reviews. PloS One 8, e59202.

Tyanova, S., Temu, T., Cox, J., 2016. The MaxQuant computational platform for mass spectrometry-based shotgun proteomics. Nat. Protoc. 11, 2301–2319. https://doi.org/10.1038/nprot.2016.136

Voß, H., Schlumbohm, S., Barwikowski, P., Wurlitzer, M., Dottermusch, M., Neumann, P., Schlüter, H., Neumann, J.E., Krisp, C., 2022. HarmonizR enables data harmonization across independent proteomic datasets with appropriate handling of missing values. Nat. Commun. 13, 3523. https://doi.org/10.1038/s41467-022-31007-x

Wang, M., Jiang, L., Jian, R., Chan, J.Y., Liu, Q., Snyder, M.P., Tang, H., 2021. RobNorm: model-based robust normalization method for labeled quantitative mass spectrometry proteomics data. Bioinformatics 37, 815–821. https://doi.org/10.1093/bioinformatics/btaa904

Wang, X., Kang, D.D., Shen, K., Song, C., Lu, S., Chang, L.-C., Liao, S.G., Huo, Z., Tang, S., Ding, Y., Kaminski, N., Sibille, E., Lin, Y., Li, J., Tseng, G.C., 2012. An R package suite for microarray meta-analysis in quality control, differentially expressed gene analysis and pathway enrichment detection. Bioinformatics 28, 2534–2536. https://doi.org/10.1093/bioinformatics/bts485

Whitlock, M.C., 2005. Combining probability from independent tests: the weighted Z-method is superior to Fisher's approach. J. Evol. Biol. 18, 1368–1373. https://doi.org/10.1111/j.1420-9101.2005.00917.x

Wiśniewski, J.R., 2018. Filter-Aided Sample Preparation for Proteome Analysis, in: Becher, D. (Ed.), Microbial Proteomics: Methods and Protocols. Springer, New York, NY, pp. 3–10. https://doi.org/10.1007/978-1-4939-8695-8_1

Xu, W., Comhair, S.A.A., Chen, R., Hu, B., Hou, Y., Zhou, Y., Mavrakis, L.A., Janocha, A.J., Li, L., Zhang, D., Willard, B.B., Asosingh, K., Cheng, F., Erzurum, S.C., 2019. Integrative proteomics and phosphoproteomics in pulmonary arterial hypertension. Sci. Rep. 9, 18623. https://doi.org/10.1038/s41598-019-55053-6

Zhang, Y., Chen, F., Chandrashekar, D.S., Varambally, S., Creighton, C.J., 2022. Proteogenomic characterization of 2002 human cancers reveals pan-cancer molecular subtypes and associated pathways. Nat. Commun. 13, 2669. https://doi.org/10.1038/s41467-022-30342-3

Zhu, Y., Orre, L.M., Zhou Tran, Y., Mermelekas, G., Johansson, H.J., Malyutina, A., Anders, S., Lehtiö, J., 2020. DEqMS: A Method for Accurate Variance Estimation in Differential Protein Expression Analysis. Mol. Cell. Proteomics MCP 19, 1047–1057. https://doi.org/10.1074/mcp.TIR119.001646

Zolotareva, O., Nasirigerdeh, R., Matschinske, J., Torkzadehmahani, R., Bakhtiari, M., Frisch, T., Späth, J., Blumenthal, D.B., Abbasinejad, A., Tieri, P., Kaissis, G., Rückert, D., Wenke, N.K., List, M., Baumbach, J., 2021. Flimma: a federated and privacy-aware tool for differential gene expression analysis. Genome Biol. 22, 338. https://doi.org/10.1186/s13059-021-02553-2

# Supplementary methods

## Design mask creation (FedProt step)

Each client has a design matrix $X^i$ with dimensions $m^i \times v$ and an intensity matrix $Y^i$ with dimensions $n \times m^i$, where $n$ is the number of protein groups shared between clients, $v$ is the number of variables in the design, and $m^i$, is the number of samples. The binary mask matrix $D$ has the number of rows equal to protein groups $n$, and columns corresponding to columns in the design matrix (see Figure M1)

On the first step, local masks $D^i$ are generated by each client (Figure M1-1). Clients locally check their data and put "1" into a local mask's cell if either they do not have data for any of the protein groups (all values in $Y^i_{p,\cdot}$ are NA) or they have only "0" in the design for a column (all values in $X^i_{\cdot,s}$ are 0):

$$\mathrm{d}^i_{p,s} = \begin{cases} 1 & \text{if } \forall j\ (Y^i_{p,j} \text{ is NA})\ \vee\ \forall j\ (X^i_{j,s} = 0) \\ 0 & \text{otherwise} \end{cases},$$

where $d^i_{p,s}$ is an element of the local mask $\mathrm{D}^i$ for i-th client, $p = 1, \ldots, n$ indexes the protein groups, $s = 1, \ldots, v$ indexes the variables in the design, and $j = 1, \ldots, m_i$ indexes the samples.

After local clients' masks are aggregated, the Coordinator checks the number of “1” for each column-row combination and if this number is equal to the number of clients ($k$), then the element $d_{p,s}$ of the global mask $D$ has “1” (True) for this value (Figure M1-2). Mathematically, this is represented as:

$$\mathrm{d}_{p,s} = \begin{cases} 1 & \text{if } \sum_{i=1}^{k} \mathrm{d}^i_{p,s} = k \\ 0 & \text{otherwise} \end{cases}.$$

This aggregated mask is sent back to the clients, and only the reference client updates it (Figure M1-3). If a reference client doesn’t have data for a specific protein (row), the last present client becomes the reference instead, and is excluded from the design by replacing “0” (False) with “1” (True) in the mask. This ensures that if all data for a protein group are NA, the last zero within the relevant batch (TMT batch, cohort) columns is set to one, ensuring the proper exclusion of the protein group from the design and subsequent calculations.

Aggregation is performed similar to the step 2 (Figure M1-4):

$$\mathrm{d}_{p,s} = \begin{cases} 1 & \text{if } \sum_{i=1}^{k} \mathrm{d}^i_{p,s} > 0, \\ 0 & \text{otherwise} \end{cases}.$$

Protein group (row) True columns will be excluded from the design and subsequent calculations. The mask is created only for binary variables in the design (cohorts, target classes). If categorical variables need to be analyzed, we suggest transforming them to binary before analysis.

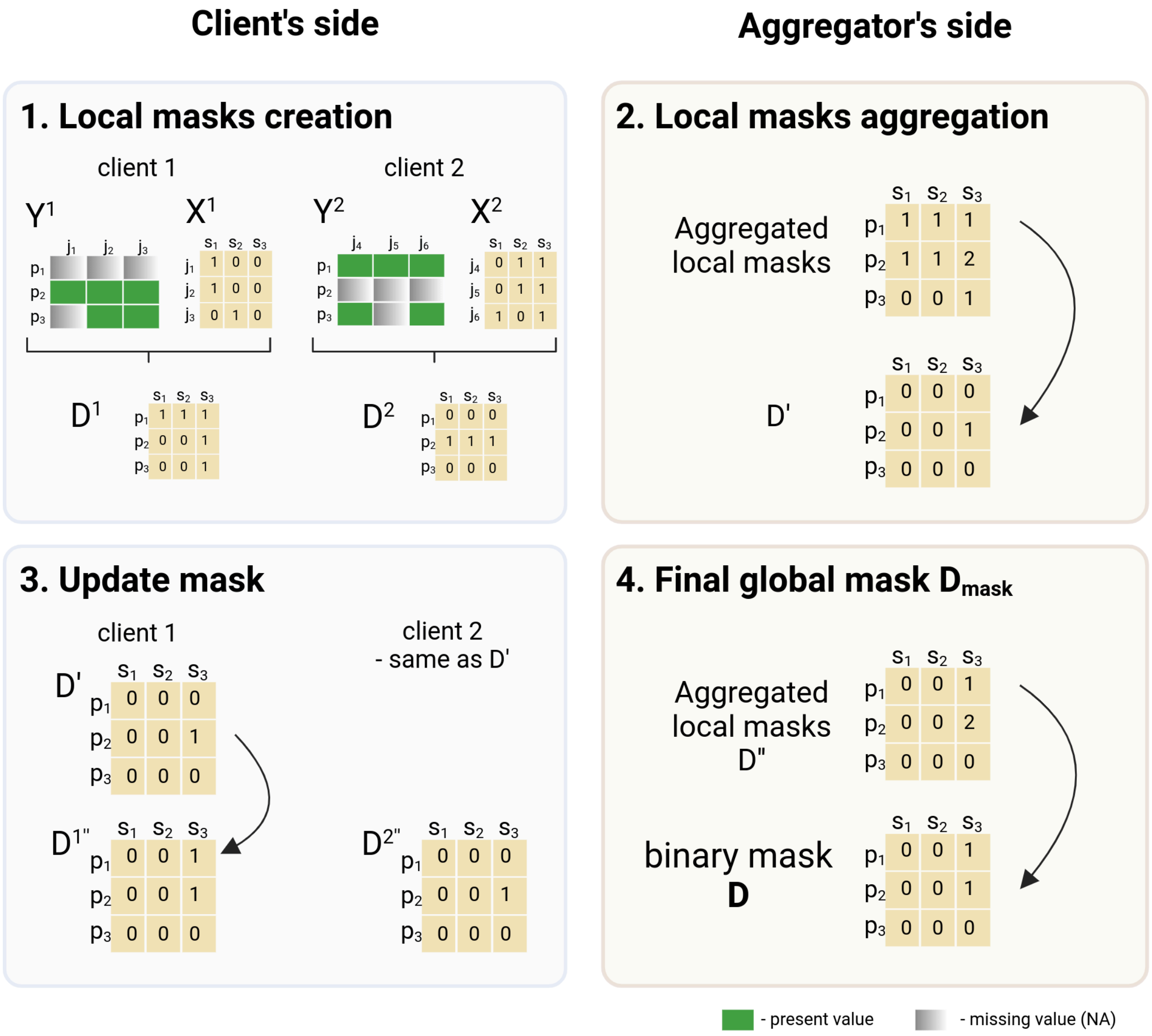

**Figure M1.**

Visual representation of binary mask $D$ creation process in FedProt.

To facilitate visualization, the example is drawn for 2 clients. However, the minimum number of clients in FedProt is 3. The formulas and description are given in the text. Figure was created with BioRender.com.

# Supplementary

## Figures

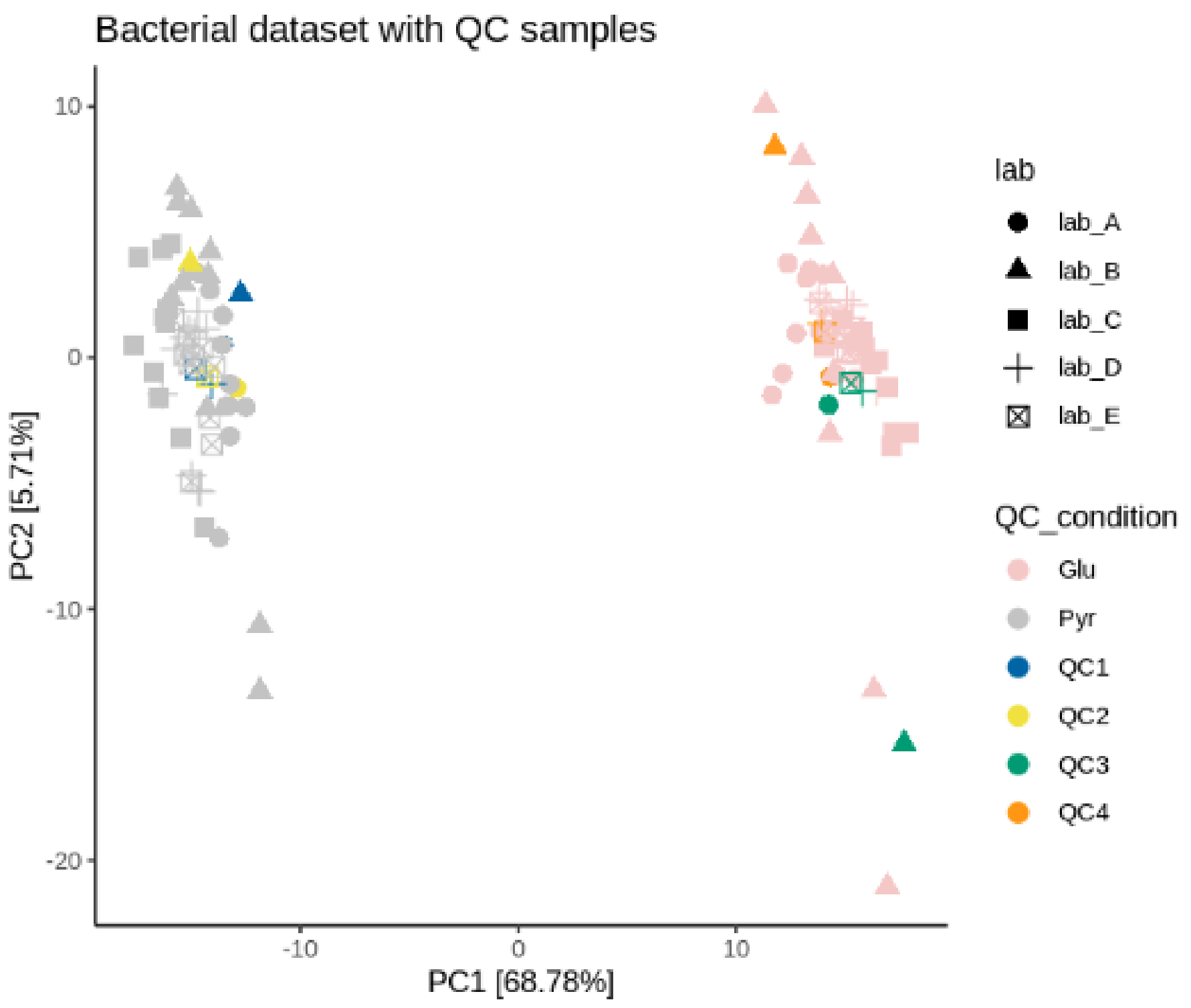

**Figure S1.**
PCA plot for bacterial dataset including QC samples after batch effect correction using limma removeBatchEffect function. These samples represent technical replicates generated for QC purposes and were removed from the dataset during FedProt evaluation.

A

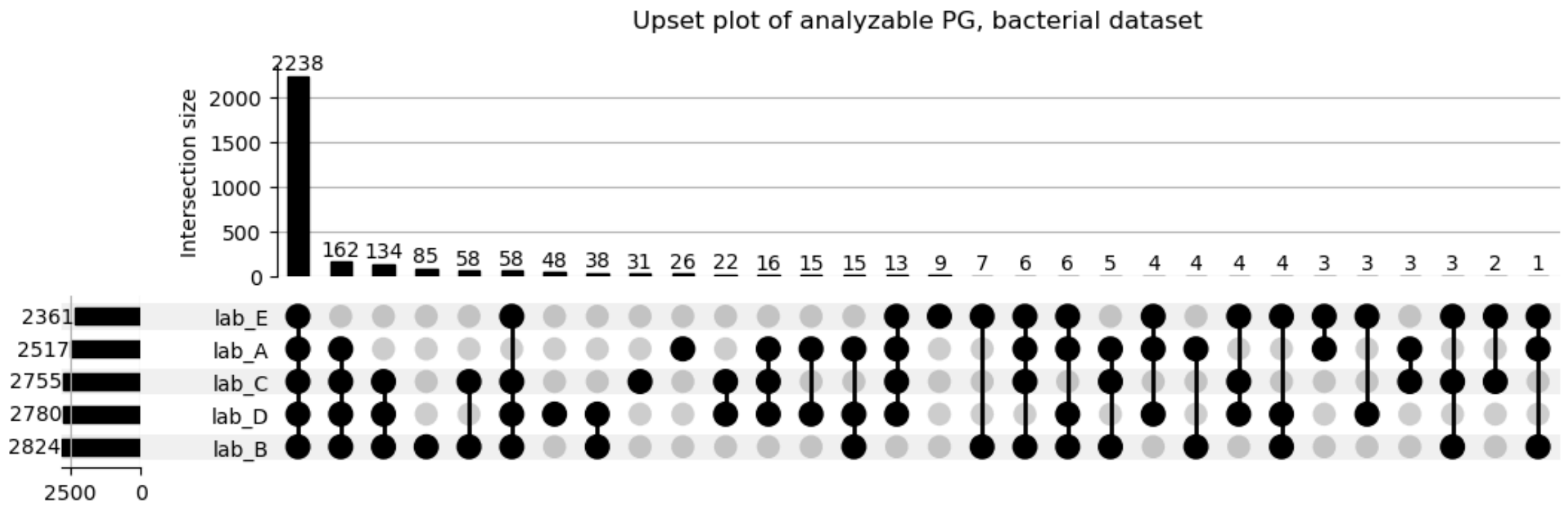

B

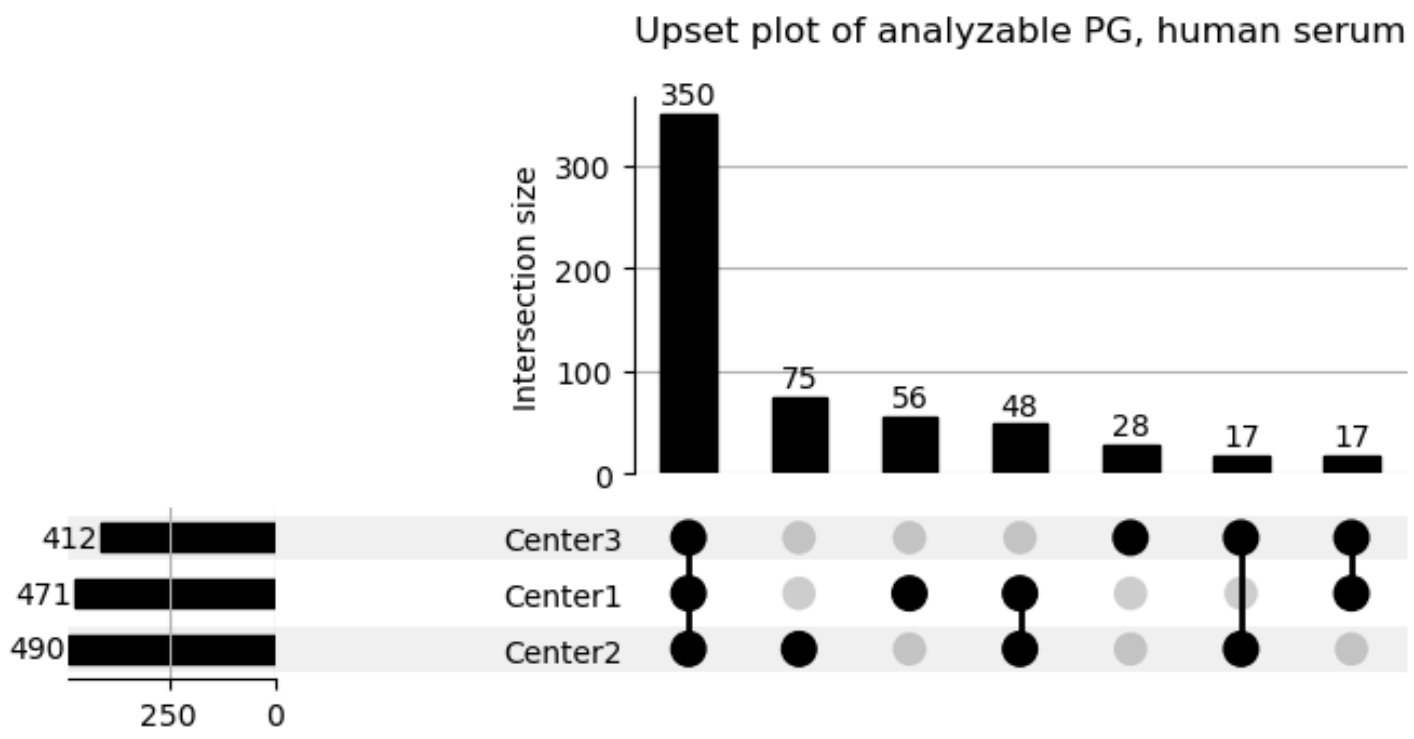

**Figure S2**.

The number of protein groups that could be analyzed by the DEqMS method inside each lab separately. A — for the bacterial dataset; B — for the human plasma dataset.

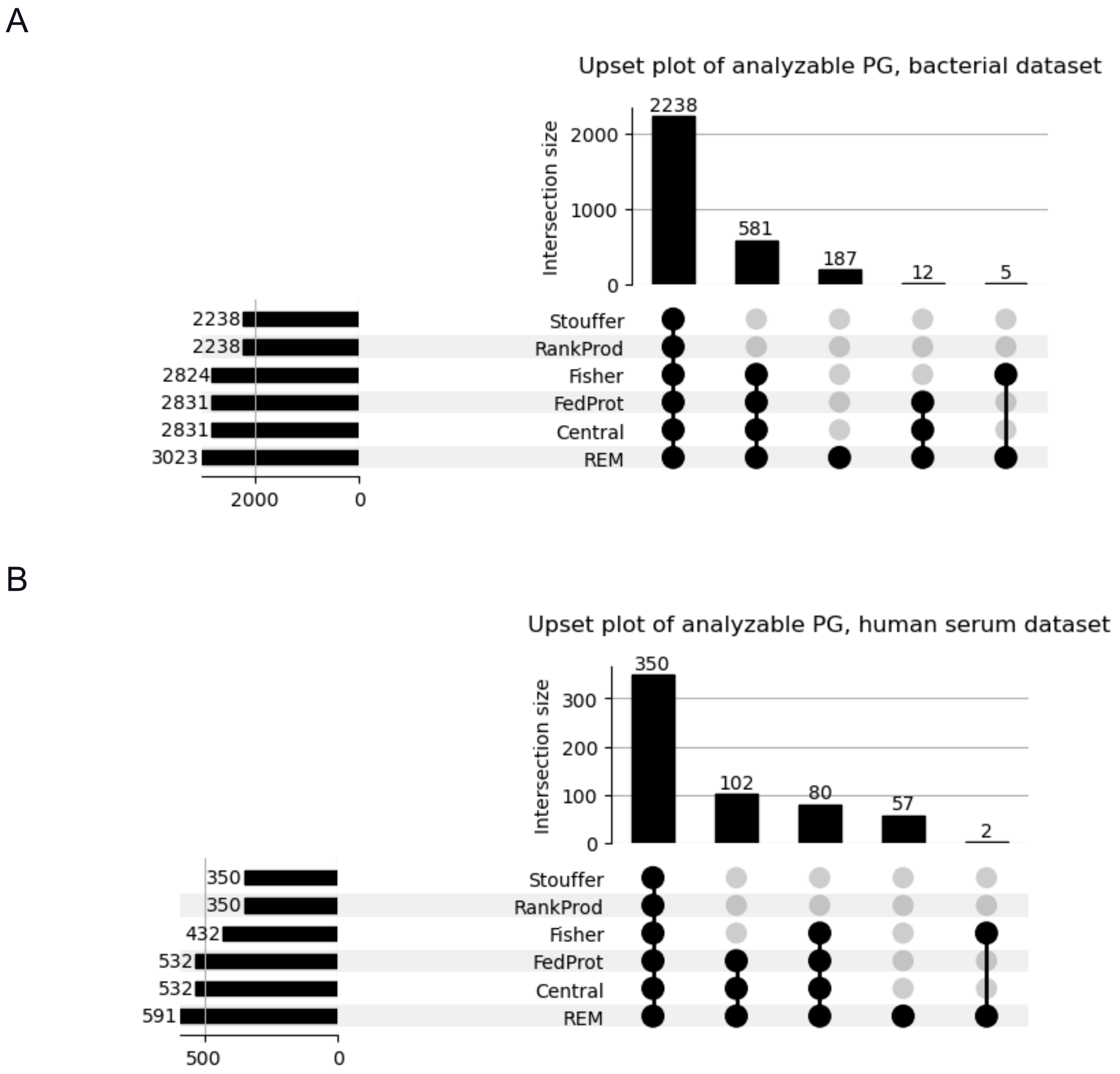

**Figure S3**.

The number of protein groups (PG) that could be analyzed by central DEqMS method, FedProt and meta-analysis methods. A — for the bacterial dataset; B — for the human serum dataset.

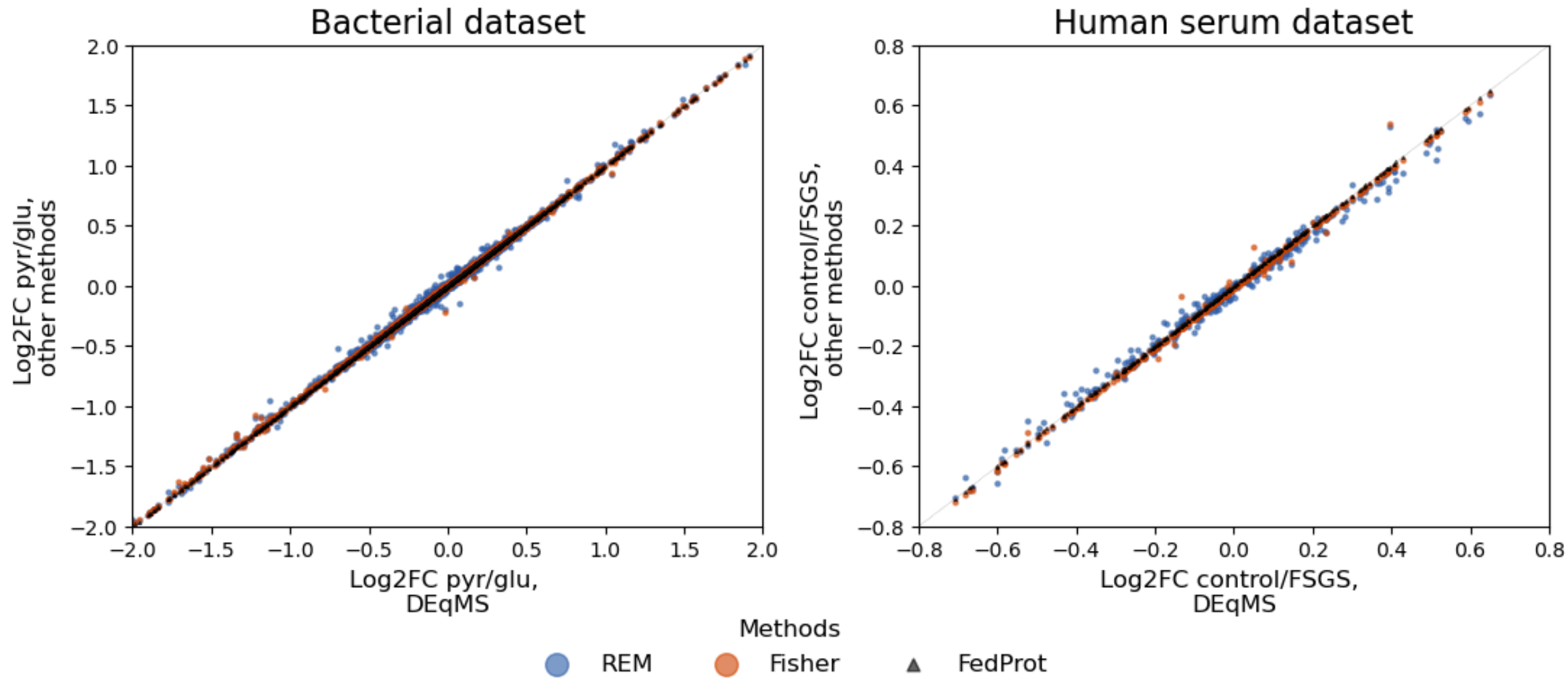

**Figure S4.**
The comparison of log-fold changes computed by FedProt or meta-analysis methods (y-axis) with centralized analysis (x-axis).
For the bacterial dataset, only values falling within the interval [-2,2] are shown; of the entire dataset, 3% of the values do not fall within this interval. For the human plasma dataset, only values falling within the interval [-0.8,0.8] are shown; of the entire dataset, 0.8% of the values do not fall within this interval.
The thin black line is the diagonal.

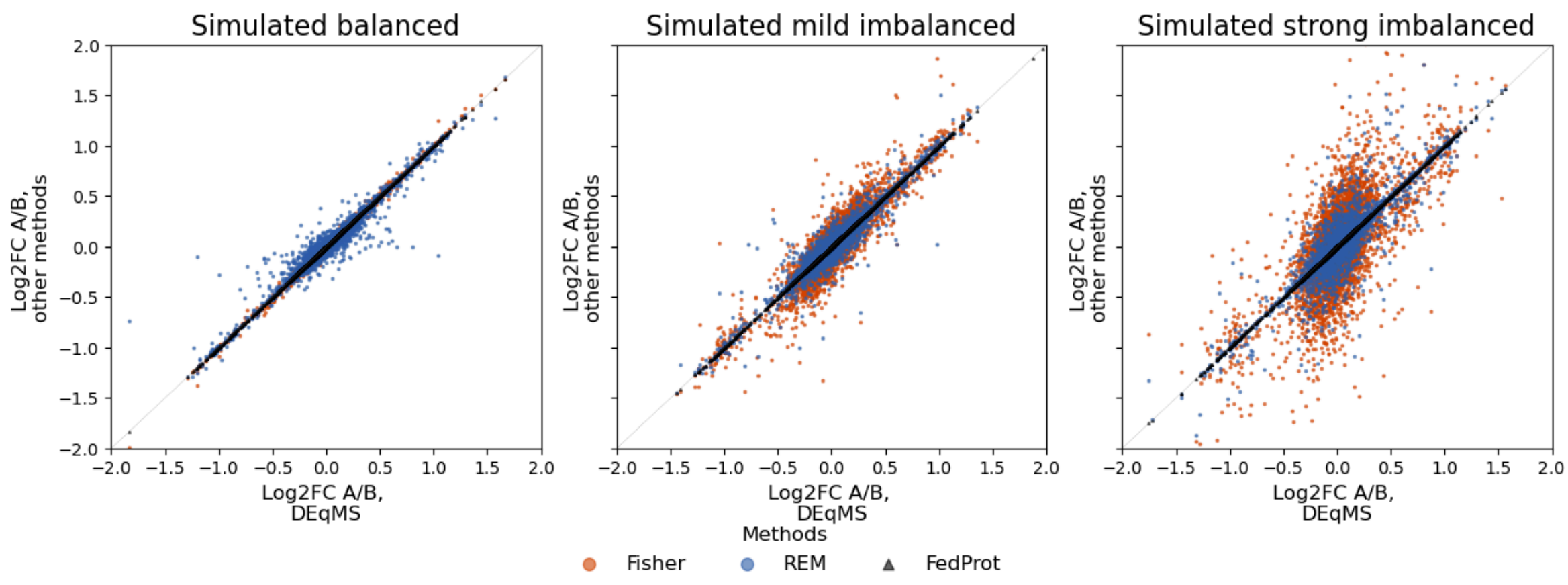

**Figure S5.**
The comparison of log-fold changes computed by FedProt or meta-analysis methods (y-axis) with centralized analysis (x-axis) for one out of 50 analysis runs for each scenario. For the simulated datasets, only values falling within the interval [-2,2] are shown. For mild imbalanced dataset, only 0.1% of values do not fall within this interval; for strong imbalanced – 0.2%.

A

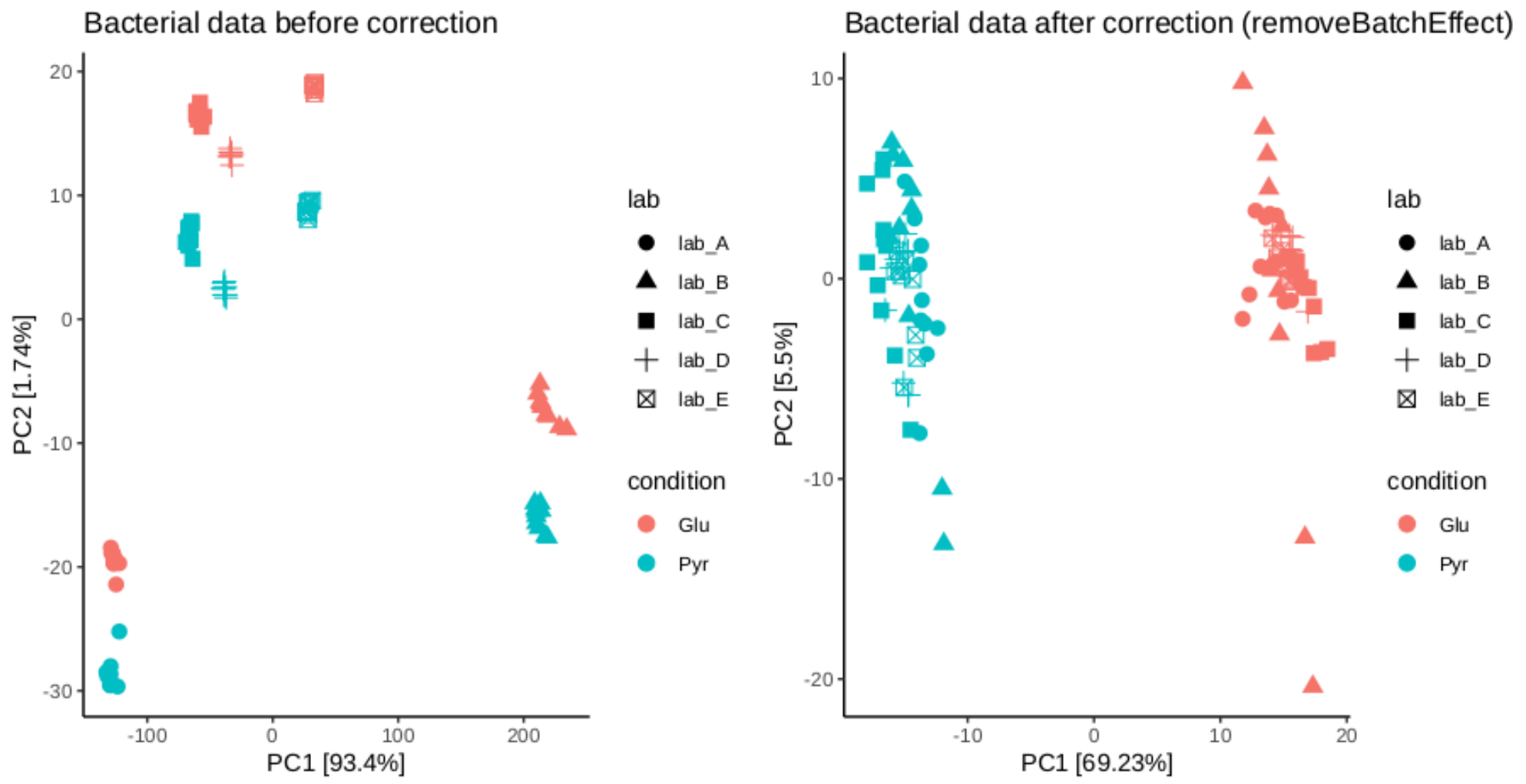

B

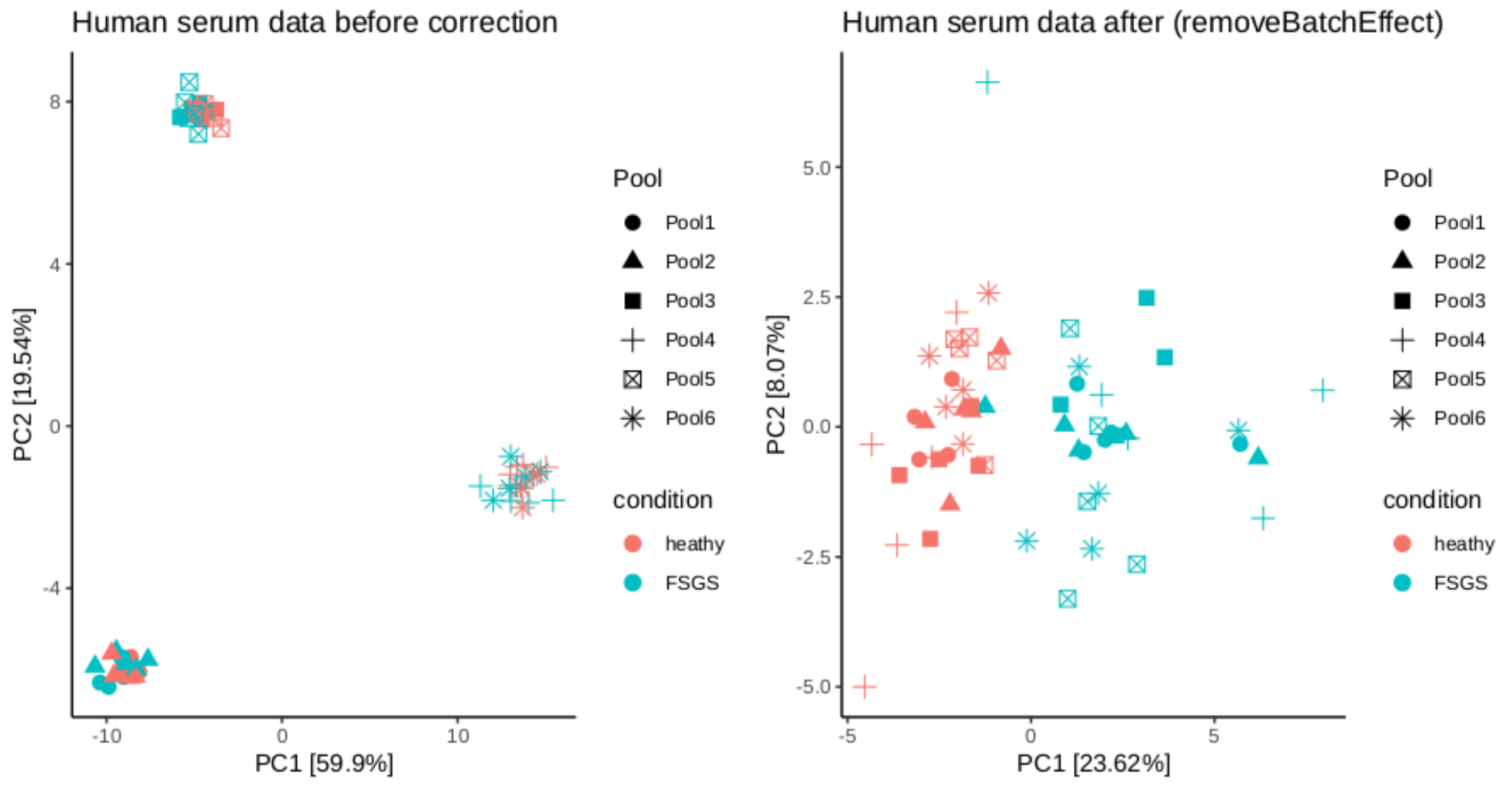

**Figure S6.**
Principal component analysis plots for all samples of bacterial (A) and human serum datasets (B) before and after batch effects correction using removeBatchEffect from limma R package[1]. The datasets were preprocessed as described in the Methods.
For the bacterial dataset, samples measured by labs A and B (on the bottom on the right and on the left) were obtained from cell pellets, the other three labs (labs C, D, and E) worked with cell lysates prepared at the lab C.

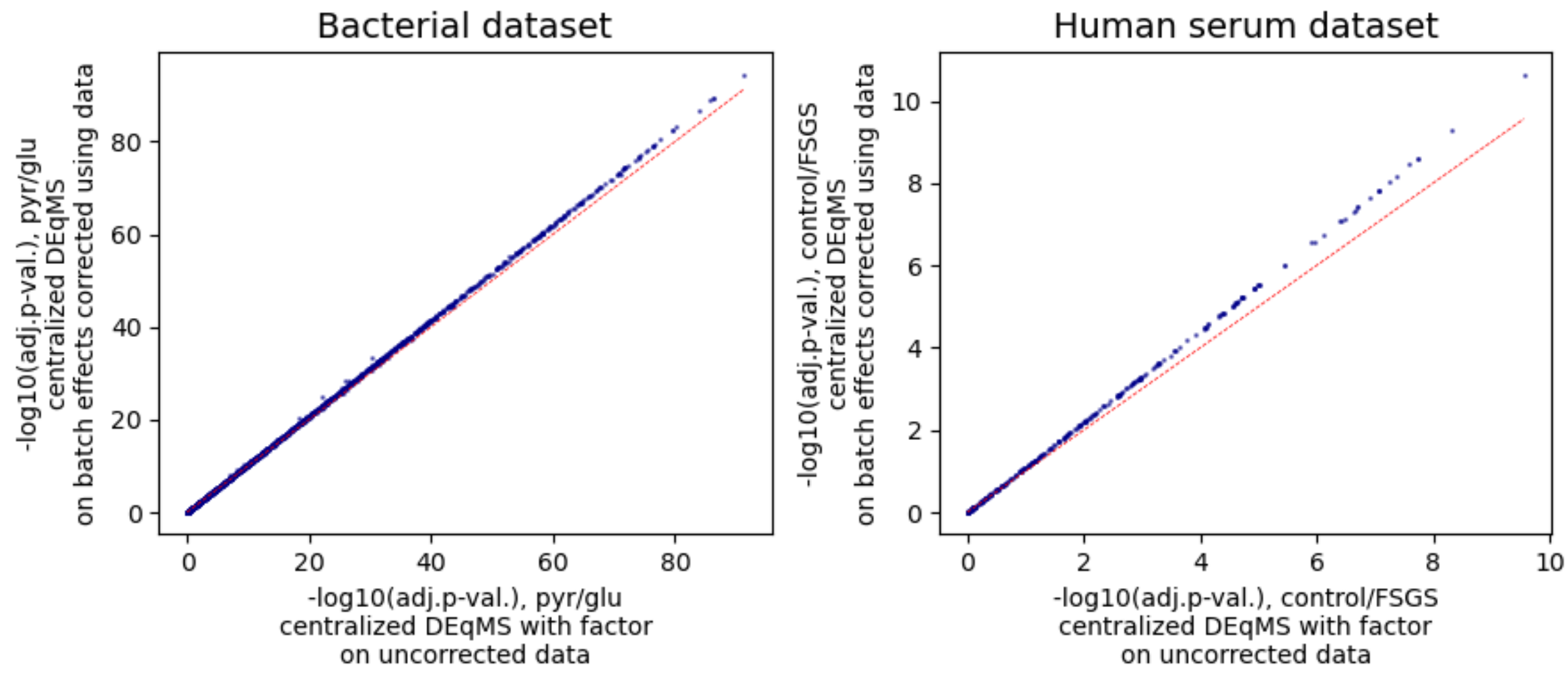

**Figure S7.**

The correlation for negative log-transformed count adjusted p-values between results on aggregated data with batch effect correction (on Y axis) and without batch effects correction, but with batch effects included in the model (on X axis).

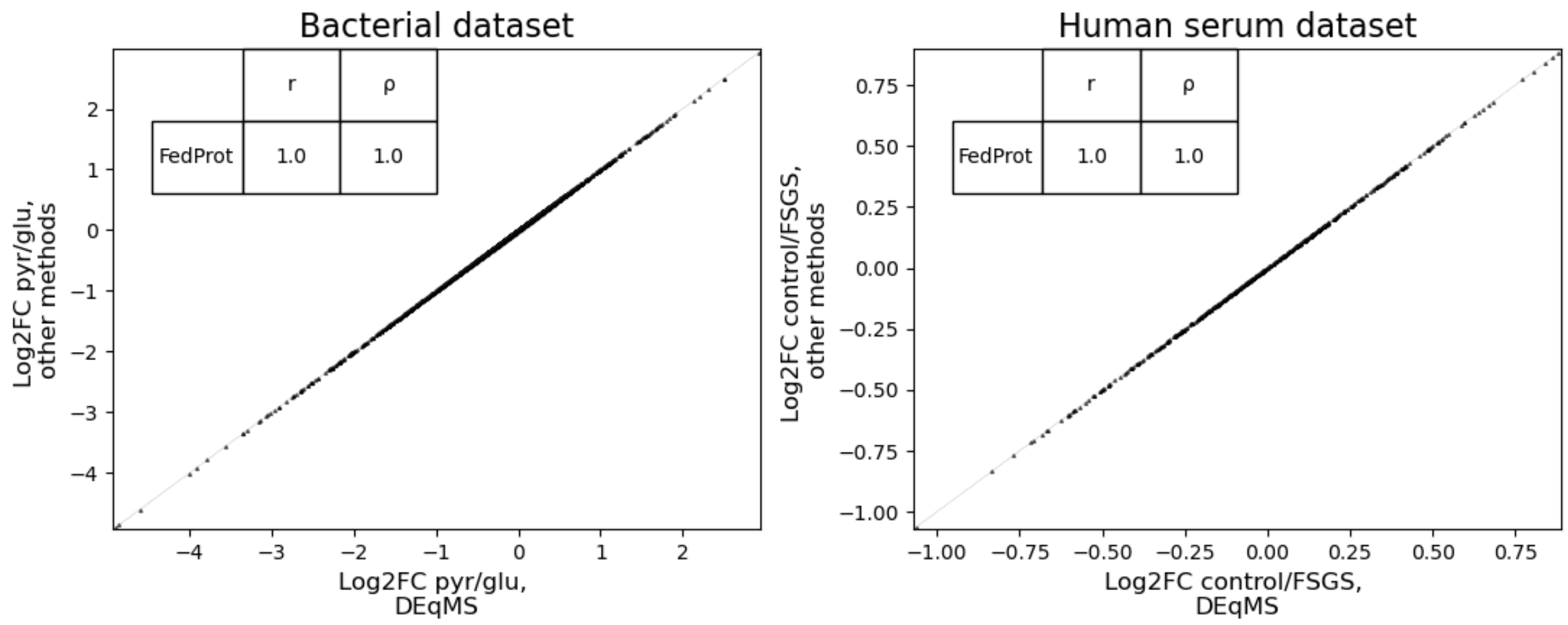

**Figure S8.**

The correlation for log-fold changes between central results on data after batch effect correction using removeBatchEffect (on Y axis) and decentralized FedProt analysis (on X axis).

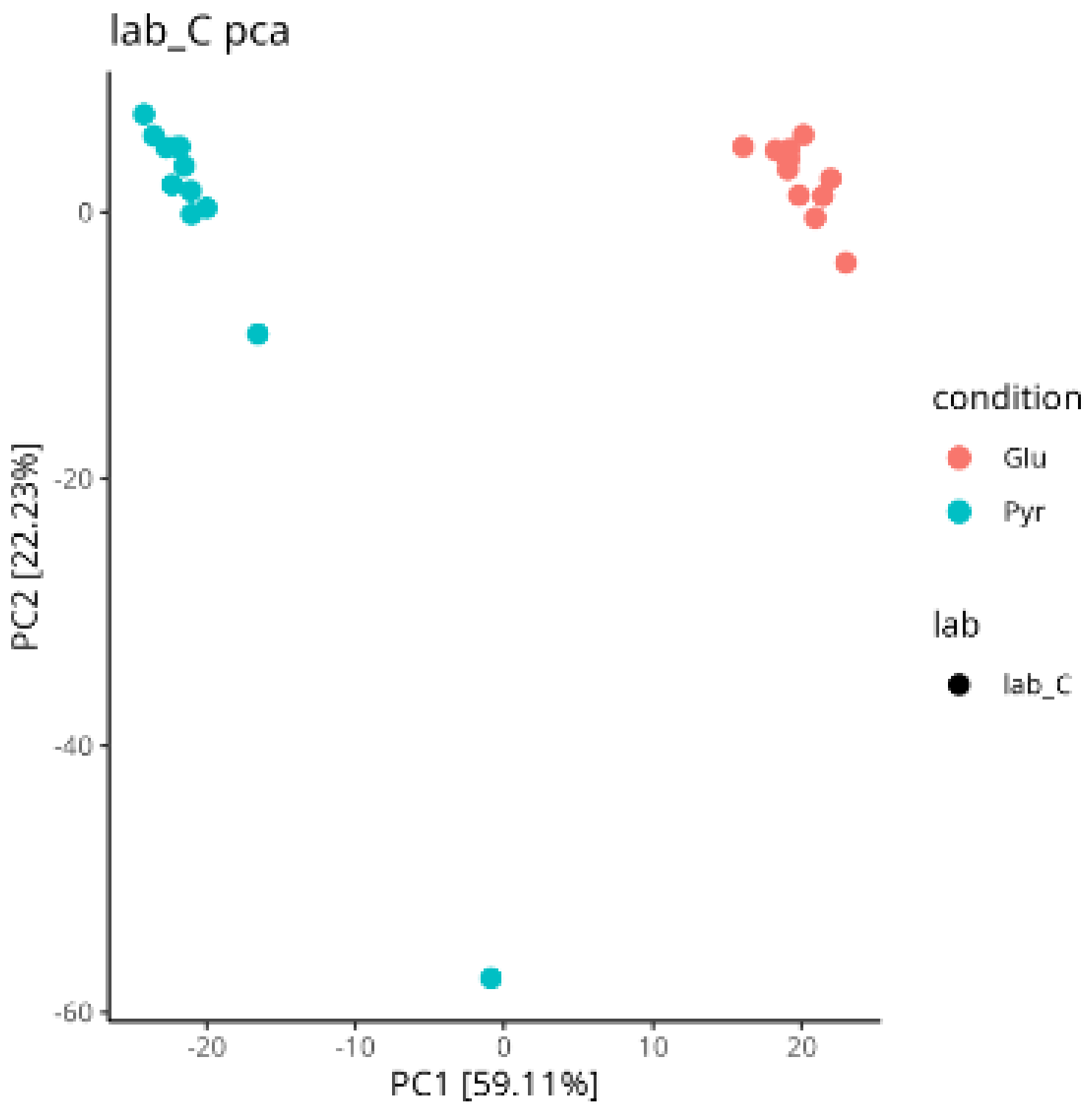

**Figure 9.**
Principal component analysis plot for data from lab C with sample excluded after quality control step.

# Tables

**Table S1.** The mean and maximum absolute differences between the log-fold change results of centralized *DEqMS* and FedProt or selected meta-analysis approaches. The lowest differences are shown in bold font. The results for the Fisher's, Stouffer's, and for the RankProd methods are the same because they use the same approach to estimate log-fold-changes.

| Dataset | **Method** | **Mean difference** | **Maximal difference** |
|---|---|---|---|
| **Bacterial** | **FedProt** | **9.81E-15** | **5.15E-14** |
| | **Fisher** | 0.006 | 0.194 |
| | **REM** | 0.017 | 0.222 |
| **Human serum** | **FedProt** | **9.07E-15** | **3.54E-14** |
| | **Fisher** | 0.010 | 0.146 |
| | **REM** | 0.019 | 0.133 |

**Table S2.** The mean and maximum absolute differences between log-fold changes of centralized *DEqMS* and FedProt or selected meta-analysis approaches. The lowest differences are shown in bold font. The generation of simulated data and the subsequent data analysis were repeated 50 times — mean and standard deviation for the mean absolute differences for these analyses results are provided.

| **Dataset** | **Method** | **Mean difference** | **Maximal difference** |
|---|---|---|---|
| **Simulated, balanced** | **FedProt** | **6.62E-16 ± 8.60E-18** | **8.04E-15 ± 1.70E-15** |
| | **Fisher** | 0.01 ± 1.51E-04 | 0.29 ± 0.23 |
| | **REM** | 0.03 ± 6.43E-04 | 1.85 ± 1.25 |
| **Simulated, mild imbalance** | **FedProt** | **1.18E-15 ± 1.88E-17** | **1.42E-14 ± 2.25E-15** |
| | **Fisher** | 0.09 ± 0.001 | 1.74 ± 0.83 |
| | **REM** | 0.04 ± 0.001 | 1.44 ± 0.57 |
| **Simulated, strong imbalance** | **FedProt** | **2.50E-15 ± 3.86E-17** | **3.26E-14 ± 6.02E-15** |
| | **Fisher** | 0.22 ± 0.003 | 5.27 ± 5.55 |
| | **REM** | 0.06 ± 0.002 | 2.41 ± 0.91 |

**Table S3**. Variable window scheme for data-independent acquisition in Lab C. z = charge state, m/z = window center, Isolation window (m/z) = window width.

| centered m/z | z | Isolation Window width (m/z) |
|---|---|---|
| 375 | 2 | 30 |
| 399 | 2 | 20 |
| 413.5 | 2 | 11 |
| 423.5 | 2 | 11 |
| 433.5 | 2 | 11 |
| 443.5 | 2 | 11 |
| 453.5 | 2 | 11 |
| 463.5 | 2 | 11 |
| 473.5 | 2 | 11 |
| 483.5 | 2 | 11 |
| 493.5 | 2 | 11 |
| 503.5 | 2 | 11 |
| 513.5 | 2 | 11 |
| 523.5 | 2 | 11 |
| 533.5 | 2 | 11 |
| 543.5 | 2 | 11 |
| 553.5 | 2 | 11 |
| 563.5 | 2 | 11 |
| 573.5 | 2 | 11 |
| 583.5 | 2 | 11 |
| 593.5 | 2 | 11 |
| 603.5 | 2 | 11 |
| 613.5 | 2 | 11 |
| 623.5 | 2 | 11 |
| 635.5 | 2 | 15 |
| 649.5 | 2 | 15 |
| 663.5 | 2 | 15 |
| 677.5 | 2 | 15 |
| 691.5 | 2 | 15 |
| 705.5 | 2 | 15 |
| 722 | 2 | 20 |
| 741 | 2 | 20 |
| 760 | 2 | 20 |
| 781.5 | 2 | 25 |
| 805.5 | 2 | 25 |
| 834.5 | 2 | 35 |
| 871 | 2 | 40 |
| 920 | 2 | 60 |
| 991.5 | 2 | 85 |
| 1166.5 | 2 | 267 |

**Table S4**. DIA-NN v1.8.1 run parameters for the bacterial dataset.

| |
|---|
| Output will be filtered at 0.01 FDR |
| Precursor/protein x samples expression level matrices will be saved along with the main report |
| A spectral library will be generated |
| Deep learning will be used to generate a new in silico spectral library from peptides provided |
| Library-free search enabled |
| Min fragment m/z set to 200 |
| Max fragment m/z set to 1800 |
| N-terminal methionine excision enabled |
| In silico digest will involve cuts at K*,R* |
| Maximum number of missed cleavages set to 2 |
| Min peptide length set to 7 |
| Max peptide length set to 30 |
| Min precursor m/z set to 360 |
| Max precursor m/z set to 1800 |
| Min precursor charge set to 1 |
| Max precursor charge set to 4 |
| Cysteine carbamidomethylation enabled as a fixed modification |
| Maximum number of variable modifications set to 1 |
| Modification UniMod:35 with mass delta 15.9949 at M will be considered as variable |
| Modification UniMod:1 with mass delta 42.0106 at *n will be considered as variable |
| A spectral library will be created from the DIA runs and used to reanalyse them; .quant files will only be saved to disk during the first step |
| When generating a spectral library, in silico predicted spectra will be retained if deemed more reliable than experimental ones |
| DIA-NN will optimise the mass accuracy automatically using the first run in the experiment. This is useful primarily for quick initial analyses, when it is not yet known which mass accuracy setting works best for a particular acquisition scheme. |
| The following variable modifications will be scored: UniMod:1 |

# Supplementary references

1. Ritchie, M. E. *et al.* limma powers differential expression analyses for RNA-sequencing and microarray studies. *Nucleic Acids Res.* **43**, e47 (2015).